\documentclass[%
 prx,
 jmp,%
 amsmath,amssymb,
 floatfix,
reprint,%
 showpacs,%
]{revtex4-2}

\usepackage{comment}
\usepackage{amssymb}
\usepackage{amsmath}
\usepackage{mathtools}
\usepackage{relsize}
\usepackage{array}
\usepackage{bm}
\usepackage{graphicx}
\usepackage{bigints}
\usepackage{ifthen}
\usepackage{multirow}
\usepackage{threeparttable}
\usepackage{rotating}
\usepackage{hyperref}
\usepackage{xcolor}
\usepackage{empheq}
\usepackage[normalem]{ulem}
\usepackage{media9}

\usepackage[draft,footnote,marginclue,innerlayout=inline]{fixme}
\fxloadlayouts{marginnote}
\fxsetup{theme=color,mode=multiuser}
\FXRegisterAuthor{V}{anV}{{\color{red}{VG}}}

\DeclareGraphicsExtensions{.pdf,.png,.eps}
\graphicspath{{figs/}{./}}
\usepackage{cleveref}
\crefname{equation}{}{}
\Crefname{equation}{}{}
\crefname{figure}{Fig.~\!\!\!}{Figs.~\!\!\!}
\Crefname{figure}{Fig.~\!\!\!}{Figs.~\!\!\!}

\usepackage{tikz}
\usetikzlibrary{intersections,arrows,decorations,decorations.pathmorphing,calc}
\usepackage{pgfplots}
\pgfplotsset{compat=newest}
\pgfplotsset{plot coordinates/math parser=false}

\makeatletter
\let\jnl@style=\rmfamily
\def\ref@jnl#1{{\jnl@style#1}}
\@ifundefined{aap}{\newcommand\aap{\ref@jnl{A\&A}}}{}%
\@ifundefined{apj}{\newcommand\apj{\ref@jnl{ApJ}}}{}%
\@ifundefined{apjl}{\newcommand\apjl{\ref@jnl{Astrophys.~J.~Lett.}}}{}%
\@ifundefined{ssr}{\newcommand\ssr{\ref@jnl{Space~Sci.~Rev.}}}{}%
\@ifundefined{jgr}{\newcommand\jgr{\ref@jnl{J.~Geophys.~Res.}}}{}%
\@ifundefined{jatp}{\newcommand\jatp{\ref@jnl{J.~Atmos.~Terr.~Phys.}}}{}%
\@ifundefined{solphys}{\newcommand\solphys{\ref@jnl{Sol.~Phys.}}}{}%
\@ifundefined{pp}{\newcommand\pp{\ref@jnl{Phys.~Plasmas}}}{}%
\@ifundefined{mnras}{\newcommand\mnras{\ref@jnl{MNRAS}}}{}%
\@ifundefined{apjs}{\newcommand\apjs{\ref@jnl{ApJS}}}{}%
\@ifundefined{azh}{\newcommand\azh{\ref@jnl{AZh}}}{}%
\catcode`\@=11

\def\beq#1\eeq{\begin{equation}#1\end{equation}}
\def\bes#1\ees{\begin{subequations}#1\end{subequations}}
\def\bea#1\eea{\begin{align}#1\end{align}}
\def\n{\nonumber\\}

\newcommand{\D}[2]{\frac{\partial #1}{\partial #2}}
\newcommand{\dd}[2]{\frac{d #1}{d #2}}
\newcommand{\mvec}[1]{\bm{#1}}

\usepackage{scalerel}

\newcommand{\DefRef}[2]{%
  \expandafter\newcommand\csname ref-#1\endcsname{#2}%
}
\newcommand{\Ref}[1]{\csname ref-#1\endcsname}
\DefRef{S01c0}{https://www.dropbox.com/sh/aads9hluqqzqfi6/AAAD0AVf2J4gRR4vuQriGuCea/S04-2.mp4?dl=0}
\DefRef{S01c1}{https://www.dropbox.com/sh/aads9hluqqzqfi6/AAAME1BQuUYX3-c3VUHhx4QPa/S04-1.mp4?dl=0}
\DefRef{S02c0}{https://www.dropbox.com/sh/aads9hluqqzqfi6/AAAhDzdznVK13ElTBxxWLO_0a/S06-1.mp4?dl=0}
\DefRef{S02c1}{https://www.dropbox.com/sh/aads9hluqqzqfi6/AAA3q5apJJTs1iq7OV2msjEwa/S06-2.mp4?dl=0}
\DefRef{S03c0}{https://www.dropbox.com/sh/aads9hluqqzqfi6/AABL5oJorExyRmN4lFY1ygPIa/S01-1.mp4?dl=0}
\DefRef{S03c1}{https://www.dropbox.com/sh/aads9hluqqzqfi6/AAA3tylPGcQVFHvQRUdaGAAfa/S01-2.mp4?dl=0}
\DefRef{S04c0}{https://www.dropbox.com/sh/aads9hluqqzqfi6/AABFWWRwute9R7-bmmzlQMnya/S02-1.mp4?dl=0}
\DefRef{S04c1}{https://www.dropbox.com/sh/aads9hluqqzqfi6/AADYgZ6x5VWLI2qlB6PKwBJea/S02-2.mp4?dl=0}
\DefRef{S05c0}{https://www.dropbox.com/sh/aads9hluqqzqfi6/AACjDA4F77nHY_bEpXtSth39a/B00-3.mp4?dl=0}
\DefRef{S05c1}{https://www.dropbox.com/sh/aads9hluqqzqfi6/AACQpgjK66ogYyzqUmVU_z0oa/B00-1.mp4?dl=0}
\DefRef{S05c2}{https://www.dropbox.com/sh/aads9hluqqzqfi6/AABSkW1jFzyBvxv3cmx36KK5a/B00-2.mp4?dl=0}
\DefRef{S06c0}{https://www.dropbox.com/sh/aads9hluqqzqfi6/AABXhFNmRW5SFqGhEDpMX6K9a/B01-3.mp4?dl=0}
\DefRef{S06c1}{https://www.dropbox.com/sh/aads9hluqqzqfi6/AADddg9sVxabkpq9P-bHU3R1a/B01-1.mp4?dl=0}
\DefRef{S06c2}{https://www.dropbox.com/sh/aads9hluqqzqfi6/AAD_Vq9D5iGCXGnVFW4URumBa/B01-2.mp4?dl=0}
\DefRef{S07c0}{https://www.dropbox.com/sh/aads9hluqqzqfi6/AABgQuRgR4N4JbzcRDuO2hPIa/B03-3.mp4?dl=0}
\DefRef{S07c1}{https://www.dropbox.com/sh/aads9hluqqzqfi6/AACg74uWizF3JjmNXMWsDR34a/B03-1.mp4?dl=0}
\DefRef{S07c2}{https://www.dropbox.com/sh/aads9hluqqzqfi6/AAC5Z3oXJgRykyeJGB3_LZJCa/B03-2.mp4?dl=0}
\DefRef{S08c0}{https://www.dropbox.com/sh/aads9hluqqzqfi6/AAA79AqpwSS-NNpj0TupYKD6a/B05-3.mp4?dl=0}
\DefRef{S08c1}{https://www.dropbox.com/sh/aads9hluqqzqfi6/AABGf0RquJz82r0f-3tJj0qoa/B05-1.mp4?dl=0}
\DefRef{S08c2}{https://www.dropbox.com/sh/aads9hluqqzqfi6/AAARyELYqUsYbZauEzlHUxQia/B05-2.mp4?dl=0}
\DefRef{S09c0}{https://www.dropbox.com/sh/aads9hluqqzqfi6/AABsn_EgyBuxnT0-gmmtIhsna/B06-3.mp4?dl=0}
\DefRef{S09c1}{https://www.dropbox.com/sh/aads9hluqqzqfi6/AAB80thJj8q166v2tAa8y9Ona/B06-1.mp4?dl=0}
\DefRef{S09c2}{https://www.dropbox.com/sh/aads9hluqqzqfi6/AABpyuvgZZfn1Sjxz3zBOmM7a/B06-2.mp4?dl=0}
\DefRef{S10c0}{https://www.dropbox.com/sh/aads9hluqqzqfi6/AAAicgl0nMXqPP5fqwiezVaTa/B07-3.mp4?dl=0}
\DefRef{S10c1}{https://www.dropbox.com/sh/aads9hluqqzqfi6/AAByd8dyJmhiiwSXn2CFkXtsa/B07-1.mp4?dl=0}
\DefRef{S10c2}{https://www.dropbox.com/sh/aads9hluqqzqfi6/AAA6dOvJguyRS_jQaoRWUi2Da/B07-2.mp4?dl=0}
\DefRef{S11c0}{https://www.dropbox.com/sh/aads9hluqqzqfi6/AAAIS8IQIKnffJzN4ajF56k5a/B08-3.mp4?dl=0}
\DefRef{S11c1}{https://www.dropbox.com/sh/aads9hluqqzqfi6/AADBMwWBIDbQK-OkgwD4Yu4Ya/B08-1.mp4?dl=0}
\DefRef{S11c2}{https://www.dropbox.com/sh/aads9hluqqzqfi6/AAA6rTbXmzttgXVH_AaRodBya/B08-2.mp4?dl=0}
\DefRef{S12c0}{https://www.dropbox.com/sh/aads9hluqqzqfi6/AABzj-VJwKBXsXOj7QEzisJHa/B09-3.mp4?dl=0}
\DefRef{S12c1}{https://www.dropbox.com/sh/aads9hluqqzqfi6/AAAitqu3x-D4t1DcPzCX2yoZa/B09-1.mp4?dl=0}
\DefRef{S12c2}{https://www.dropbox.com/sh/aads9hluqqzqfi6/AABHZ1EPrJ3XJMoV2ALSa7lra/B09-2.mp4?dl=0}
\DefRef{S13c0}{https://www.dropbox.com/sh/aads9hluqqzqfi6/AADO4mnVv5xLospYOPD2gIARa/B10-3.mp4?dl=0}
\DefRef{S13c1}{https://www.dropbox.com/sh/aads9hluqqzqfi6/AAAmpBzb15WHOsS9Igne9C6-a/B10-1.mp4?dl=0}
\DefRef{S13c2}{https://www.dropbox.com/sh/aads9hluqqzqfi6/AACF1UmEyAe4FuUxYDMwqdKUa/B10-2.mp4?dl=0}
\DefRef{S14c0}{https://www.dropbox.com/sh/aads9hluqqzqfi6/AACixcIHWiY_QC2pXlIH-uNda/B11-3.mp4?dl=0}
\DefRef{S14c1}{https://www.dropbox.com/sh/aads9hluqqzqfi6/AACYJIkuuEk07BEs0lLkelDha/B11-1.mp4?dl=0}
\DefRef{S14c2}{https://www.dropbox.com/sh/aads9hluqqzqfi6/AAAnPLF9UBfLSwQ867MdEuuja/B11-2.mp4?dl=0}
\DefRef{S15c0}{https://www.dropbox.com/sh/aads9hluqqzqfi6/AAByOxXMLj9veOtz3JdozKvXa/B12-3.mp4?dl=0}
\DefRef{S15c1}{https://www.dropbox.com/sh/aads9hluqqzqfi6/AAAhPapCGN4eg8jw78DPHRZXa/B12-1.mp4?dl=0}
\DefRef{S15c2}{https://www.dropbox.com/sh/aads9hluqqzqfi6/AAC9Ta1d9eTSjn0FyRFuRSG0a/B12-2.mp4?dl=0}
\DefRef{S16c0}{https://www.dropbox.com/sh/aads9hluqqzqfi6/AACaCzRuxllykR-Ua2Xfn7qPa/B13-3.mp4?dl=0}
\DefRef{S16c1}{https://www.dropbox.com/sh/aads9hluqqzqfi6/AAA7l-fqZqHH8NMulpeNVsFWa/B13-1.mp4?dl=0}
\DefRef{S16c2}{https://www.dropbox.com/sh/aads9hluqqzqfi6/AAD9uliqRIXkYiJiDJU5QRbia/B13-2.mp4?dl=0}
\DefRef{SA01c0}{http://abeta.ucsd.edu/Videos.git/BW/S04-2.mp4}
\DefRef{SA01c1}{http://abeta.ucsd.edu/Videos.git/BW/S04-1.mp4}
\DefRef{SA02c0}{http://abeta.ucsd.edu/Videos.git/BW/S06-1.mp4}
\DefRef{SA02c1}{http://abeta.ucsd.edu/Videos.git/BW/S06-2.mp4}
\DefRef{SA03c0}{http://abeta.ucsd.edu/Videos.git/BW/S01-1.mp4}
\DefRef{SA03c1}{http://abeta.ucsd.edu/Videos.git/BW/S01-2.mp4}
\DefRef{SA04c0}{http://abeta.ucsd.edu/Videos.git/BW/S02-1.mp4}
\DefRef{SA04c1}{http://abeta.ucsd.edu/Videos.git/BW/S02-2.mp4}
\DefRef{SA05c0}{http://abeta.ucsd.edu/Videos.git/BW/B00-3.mp4}
\DefRef{SA05c1}{http://abeta.ucsd.edu/Videos.git/BW/B00-1.mp4}
\DefRef{SA05c2}{http://abeta.ucsd.edu/Videos.git/BW/B00-2.mp4}
\DefRef{SA06c0}{http://abeta.ucsd.edu/Videos.git/BW/B01-3.mp4}
\DefRef{SA06c1}{http://abeta.ucsd.edu/Videos.git/BW/B01-1.mp4}
\DefRef{SA06c2}{http://abeta.ucsd.edu/Videos.git/BW/B01-2.mp4}
\DefRef{SA07c0}{http://abeta.ucsd.edu/Videos.git/BW/B03-3.mp4}
\DefRef{SA07c1}{http://abeta.ucsd.edu/Videos.git/BW/B03-1.mp4}
\DefRef{SA07c2}{http://abeta.ucsd.edu/Videos.git/BW/B03-2.mp4}
\DefRef{SA08c0}{http://abeta.ucsd.edu/Videos.git/BW/B05-3.mp4}
\DefRef{SA08c1}{http://abeta.ucsd.edu/Videos.git/BW/B05-1.mp4}
\DefRef{SA08c2}{http://abeta.ucsd.edu/Videos.git/BW/B05-2.mp4}
\DefRef{SA09c0}{http://abeta.ucsd.edu/Videos.git/BW/B06-3.mp4}
\DefRef{SA09c1}{http://abeta.ucsd.edu/Videos.git/BW/B06-1.mp4}
\DefRef{SA09c2}{http://abeta.ucsd.edu/Videos.git/BW/B06-2.mp4}
\DefRef{SA10c0}{http://abeta.ucsd.edu/Videos.git/BW/B07-3.mp4}
\DefRef{SA10c1}{http://abeta.ucsd.edu/Videos.git/BW/B07-1.mp4}
\DefRef{SA10c2}{http://abeta.ucsd.edu/Videos.git/BW/B07-2.mp4}
\DefRef{SA11c0}{http://abeta.ucsd.edu/Videos.git/BW/B08-3.mp4}
\DefRef{SA11c1}{http://abeta.ucsd.edu/Videos.git/BW/B08-1.mp4}
\DefRef{SA11c2}{http://abeta.ucsd.edu/Videos.git/BW/B08-2.mp4}
\DefRef{SA12c0}{http://abeta.ucsd.edu/Videos.git/BW/B09-3.mp4}
\DefRef{SA12c1}{http://abeta.ucsd.edu/Videos.git/BW/B09-1.mp4}
\DefRef{SA12c2}{http://abeta.ucsd.edu/Videos.git/BW/B09-2.mp4}
\DefRef{SA13c0}{http://abeta.ucsd.edu/Videos.git/BW/B10-3.mp4}
\DefRef{SA13c1}{http://abeta.ucsd.edu/Videos.git/BW/B10-1.mp4}
\DefRef{SA13c2}{http://abeta.ucsd.edu/Videos.git/BW/B10-2.mp4}
\DefRef{SA14c0}{http://abeta.ucsd.edu/Videos.git/BW/B11-3.mp4}
\DefRef{SA14c1}{http://abeta.ucsd.edu/Videos.git/BW/B11-1.mp4}
\DefRef{SA14c2}{http://abeta.ucsd.edu/Videos.git/BW/B11-2.mp4}
\DefRef{SA15c0}{http://abeta.ucsd.edu/Videos.git/BW/B12-3.mp4}
\DefRef{SA15c1}{http://abeta.ucsd.edu/Videos.git/BW/B12-1.mp4}
\DefRef{SA15c2}{http://abeta.ucsd.edu/Videos.git/BW/B12-2.mp4}
\DefRef{SA16c0}{http://abeta.ucsd.edu/Videos.git/BW/B13-3.mp4}
\DefRef{SA16c1}{http://abeta.ucsd.edu/Videos.git/BW/B13-1.mp4}
\DefRef{SA16c2}{http://abeta.ucsd.edu/Videos.git/BW/B13-2.mp4}
\DefRef{SSA01c0}{http://abeta.ucsd.edu/Videos.git/BW/S04-2-s.mp4}
\DefRef{SSA01c1}{http://abeta.ucsd.edu/Videos.git/BW/S04-1-s.mp4}
\DefRef{SSA02c0}{http://abeta.ucsd.edu/Videos.git/BW/S06-1-s.mp4}
\DefRef{SSA02c1}{http://abeta.ucsd.edu/Videos.git/BW/S06-2-s.mp4}
\DefRef{SSA03c0}{http://abeta.ucsd.edu/Videos.git/BW/S01-1-s.mp4}
\DefRef{SSA03c1}{http://abeta.ucsd.edu/Videos.git/BW/S01-2-s.mp4}
\DefRef{SSA04c0}{http://abeta.ucsd.edu/Videos.git/BW/S02-1-s.mp4}
\DefRef{SSA04c1}{http://abeta.ucsd.edu/Videos.git/BW/S02-2-s.mp4}
\DefRef{SSA05c0}{http://abeta.ucsd.edu/Videos.git/BW/B00-3-s.mp4}
\DefRef{SSA05c1}{http://abeta.ucsd.edu/Videos.git/BW/B00-1-s.mp4}
\DefRef{SSA05c2}{http://abeta.ucsd.edu/Videos.git/BW/B00-2-s.mp4}
\DefRef{SSA06c0}{http://abeta.ucsd.edu/Videos.git/BW/B01-3-s.mp4}
\DefRef{SSA06c1}{http://abeta.ucsd.edu/Videos.git/BW/B01-1-s.mp4}
\DefRef{SSA06c2}{http://abeta.ucsd.edu/Videos.git/BW/B01-2-s.mp4}
\DefRef{SSA07c0}{http://abeta.ucsd.edu/Videos.git/BW/B03-3-s.mp4}
\DefRef{SSA07c1}{http://abeta.ucsd.edu/Videos.git/BW/B03-1-s.mp4}
\DefRef{SSA07c2}{http://abeta.ucsd.edu/Videos.git/BW/B03-2-s.mp4}
\DefRef{SSA08c0}{http://abeta.ucsd.edu/Videos.git/BW/B05-3-s.mp4}
\DefRef{SSA08c1}{http://abeta.ucsd.edu/Videos.git/BW/B05-1-s.mp4}
\DefRef{SSA08c2}{http://abeta.ucsd.edu/Videos.git/BW/B05-2-s.mp4}
\DefRef{SSA09c0}{http://abeta.ucsd.edu/Videos.git/BW/B06-3-s.mp4}
\DefRef{SSA09c1}{http://abeta.ucsd.edu/Videos.git/BW/B06-1-s.mp4}
\DefRef{SSA09c2}{http://abeta.ucsd.edu/Videos.git/BW/B06-2-s.mp4}
\DefRef{SSA10c0}{http://abeta.ucsd.edu/Videos.git/BW/B07-3-s.mp4}
\DefRef{SSA10c1}{http://abeta.ucsd.edu/Videos.git/BW/B07-1-s.mp4}
\DefRef{SSA10c2}{http://abeta.ucsd.edu/Videos.git/BW/B07-2-s.mp4}
\DefRef{SSA11c0}{http://abeta.ucsd.edu/Videos.git/BW/B08-3-s.mp4}
\DefRef{SSA11c1}{http://abeta.ucsd.edu/Videos.git/BW/B08-1-s.mp4}
\DefRef{SSA11c2}{http://abeta.ucsd.edu/Videos.git/BW/B08-2-s.mp4}
\DefRef{SSA12c0}{http://abeta.ucsd.edu/Videos.git/BW/B09-3-s.mp4}
\DefRef{SSA12c1}{http://abeta.ucsd.edu/Videos.git/BW/B09-1-s.mp4}
\DefRef{SSA12c2}{http://abeta.ucsd.edu/Videos.git/BW/B09-2-s.mp4}
\DefRef{SSA13c0}{http://abeta.ucsd.edu/Videos.git/BW/B10-3-s.mp4}
\DefRef{SSA13c1}{http://abeta.ucsd.edu/Videos.git/BW/B10-1-s.mp4}
\DefRef{SSA13c2}{http://abeta.ucsd.edu/Videos.git/BW/B10-2-s.mp4}
\DefRef{SSA14c0}{http://abeta.ucsd.edu/Videos.git/BW/B11-3-s.mp4}
\DefRef{SSA14c1}{http://abeta.ucsd.edu/Videos.git/BW/B11-1-s.mp4}
\DefRef{SSA14c2}{http://abeta.ucsd.edu/Videos.git/BW/B11-2-s.mp4}
\DefRef{SSA15c0}{http://abeta.ucsd.edu/Videos.git/BW/B12-3-s.mp4}
\DefRef{SSA15c1}{http://abeta.ucsd.edu/Videos.git/BW/B12-1-s.mp4}
\DefRef{SSA15c2}{http://abeta.ucsd.edu/Videos.git/BW/B12-2-s.mp4}
\DefRef{SSA16c0}{http://abeta.ucsd.edu/Videos.git/BW/B13-3-s.mp4}
\DefRef{SSA16c1}{http://abeta.ucsd.edu/Videos.git/BW/B13-1-s.mp4}
\DefRef{SSA16c2}{http://abeta.ucsd.edu/Videos.git/BW/B13-2-s.mp4}
\DefRef{S01}{S04}
\DefRef{S02}{S06}
\DefRef{S03}{S01}
\DefRef{S04}{S02}
\DefRef{S05}{B00}
\DefRef{S06}{B01}
\DefRef{S07}{B03}
\DefRef{S08}{B05}
\DefRef{S09}{B06}
\DefRef{S10}{B07}
\DefRef{S11}{B08}
\DefRef{S12}{B09}
\DefRef{S13}{B10}
\DefRef{S14}{B11}
\DefRef{S15}{B12}
\DefRef{S16}{B13}
\DefRef{SS01c0}{https://www.dropbox.com/sh/aads9hluqqzqfi6/AADh0JgtHmqgghATudJec4qUa/S04-2-s.mp4?dl=0}
\DefRef{SS01c1}{https://www.dropbox.com/sh/aads9hluqqzqfi6/AAC6ZH6kaBPSdl8tRWwqBcQDa/S04-1-s.mp4?dl=0}
\DefRef{SS02c0}{https://www.dropbox.com/sh/aads9hluqqzqfi6/AABsduaPtwU-fN43hR5WHEh0a/S06-1-s.mp4?dl=0}
\DefRef{SS02c1}{https://www.dropbox.com/sh/aads9hluqqzqfi6/AAB9CgRtWItUpWjaWTqdIg28a/S06-2-s.mp4?dl=0}
\DefRef{SS03c0}{https://www.dropbox.com/sh/aads9hluqqzqfi6/AAB2mezFGiC_Gxn2PQaUmVT2a/S01-1-s.mp4?dl=0}
\DefRef{SS03c1}{https://www.dropbox.com/sh/aads9hluqqzqfi6/AAAYTALDyw09aVaSTpe5H4SJa/S01-2-s.mp4?dl=0}
\DefRef{SS04c0}{https://www.dropbox.com/sh/aads9hluqqzqfi6/AAAgZ5JtOK2VFJqAIXUnAx5ca/S02-1-s.mp4?dl=0}
\DefRef{SS04c1}{https://www.dropbox.com/sh/aads9hluqqzqfi6/AABBYZgtkrKtGQdCt31kwQYla/S02-2-s.mp4?dl=0}
\DefRef{SS05c1}{https://www.dropbox.com/sh/aads9hluqqzqfi6/AACqPiMDgMygTAtqdmZt2iv_a/B00-1-s.mp4?dl=0}
\DefRef{SS05c2}{https://www.dropbox.com/sh/aads9hluqqzqfi6/AAA_orXls1lFnw3BZFqW9ttWa/B00-2-s.mp4?dl=0}
\DefRef{SS05c0}{https://www.dropbox.com/sh/aads9hluqqzqfi6/AACgCXJvoiLf9U1KxO6ftyJFa/B00-3-s.mp4?dl=0}
\DefRef{SS06c1}{https://www.dropbox.com/sh/aads9hluqqzqfi6/AAA5adhXBq7Qi6BT_kUr7b_Pa/B01-1-s.mp4?dl=0}
\DefRef{SS06c2}{https://www.dropbox.com/sh/aads9hluqqzqfi6/AADEDuZNYsCV-EEh7qIgqWPWa/B01-2-s.mp4?dl=0}
\DefRef{SS06c0}{https://www.dropbox.com/sh/aads9hluqqzqfi6/AACOs8yKxqbhAGtOKNKRFoKta/B01-3-s.mp4?dl=0}
\DefRef{SS07c1}{https://www.dropbox.com/sh/aads9hluqqzqfi6/AADz-AGlBhQzwAJ7e4LkFr_Va/B03-1-s.mp4?dl=0}
\DefRef{SS07c2}{https://www.dropbox.com/sh/aads9hluqqzqfi6/AABxgsLcGp8PhN8br9OASYZYa/B03-2-s.mp4?dl=0}
\DefRef{SS07c0}{https://www.dropbox.com/sh/aads9hluqqzqfi6/AAA8taETW_Xi3fKY9GmmtHOBa/B03-3-s.mp4?dl=0}
\DefRef{SS08c1}{https://www.dropbox.com/sh/aads9hluqqzqfi6/AAAk0JhNho9khvKfWTAPZgFDa/B05-1-s.mp4?dl=0}
\DefRef{SS08c2}{https://www.dropbox.com/sh/aads9hluqqzqfi6/AABSbaMTgqtIsAuRe2LT3bToa/B05-2-s.mp4?dl=0}
\DefRef{SS08c0}{https://www.dropbox.com/sh/aads9hluqqzqfi6/AAB86_hTHCJB59u6zlKHv_RTa/B05-3-s.mp4?dl=0}
\DefRef{SS09c1}{https://www.dropbox.com/sh/aads9hluqqzqfi6/AADpyoejlJYjp0UMlT-naz0ta/B06-1-s.mp4?dl=0}
\DefRef{SS09c2}{https://www.dropbox.com/sh/aads9hluqqzqfi6/AAD5BURgrZWq1r9GzvlBHzyEa/B06-2-s.mp4?dl=0}
\DefRef{SS09c0}{https://www.dropbox.com/sh/aads9hluqqzqfi6/AACUFHH6bxaKAnQawClvhNmra/B06-3-s.mp4?dl=0}
\DefRef{SS10c1}{https://www.dropbox.com/sh/aads9hluqqzqfi6/AAB4HkbaOhmDEBxR-xkFFPoZa/B07-1-s.mp4?dl=0}
\DefRef{SS10c2}{https://www.dropbox.com/sh/aads9hluqqzqfi6/AAB-uV3JsKh4IvSYS4MzNSa8a/B07-2-s.mp4?dl=0}
\DefRef{SS10c0}{https://www.dropbox.com/sh/aads9hluqqzqfi6/AADAR4ceGqBpUNQ7S208870Ra/B07-3-s.mp4?dl=0}
\DefRef{SS11c1}{https://www.dropbox.com/sh/aads9hluqqzqfi6/AADBknvqy76CK9__oJN1AfGka/B08-1-s.mp4?dl=0}
\DefRef{SS11c2}{https://www.dropbox.com/sh/aads9hluqqzqfi6/AADec-aeHbsxxYm-6HGe-BMCa/B08-2-s.mp4?dl=0}
\DefRef{SS11c0}{https://www.dropbox.com/sh/aads9hluqqzqfi6/AADcBK4L_ri-0r0XmMmcd8-Ma/B08-3-s.mp4?dl=0}
\DefRef{SS12c1}{https://www.dropbox.com/sh/aads9hluqqzqfi6/AABia7Ke2RYkEVlBkm-Y3UZFa/B09-1-s.mp4?dl=0}
\DefRef{SS12c2}{https://www.dropbox.com/sh/aads9hluqqzqfi6/AAAYKJfQEbF6e4GHXVVQsGYRa/B09-2-s.mp4?dl=0}
\DefRef{SS12c0}{https://www.dropbox.com/sh/aads9hluqqzqfi6/AABVF9VQcEeZGyILBo_rqwnsa/B09-3-s.mp4?dl=0}
\DefRef{SS13c1}{https://www.dropbox.com/sh/aads9hluqqzqfi6/AAA2475HA3OdRjH-6p4SESa4a/B10-1-s.mp4?dl=0}
\DefRef{SS13c2}{https://www.dropbox.com/sh/aads9hluqqzqfi6/AABNJUR_nNfMTC033fBXKN8Ra/B10-2-s.mp4?dl=0}
\DefRef{SS13c0}{https://www.dropbox.com/sh/aads9hluqqzqfi6/AACfawadTMiINftlLpphdltwa/B10-3-s.mp4?dl=0}
\DefRef{SS14c1}{https://www.dropbox.com/sh/aads9hluqqzqfi6/AABAZmu1c7ffB76E3R0FSaOoa/B11-1-s.mp4?dl=0}
\DefRef{SS14c2}{https://www.dropbox.com/sh/aads9hluqqzqfi6/AADpHFk3DbGa510fniQ1zc5Na/B11-2-s.mp4?dl=0}
\DefRef{SS14c0}{https://www.dropbox.com/sh/aads9hluqqzqfi6/AAApUHJ2YCAHv7XOCZ8TBLRga/B11-3-s.mp4?dl=0}
\DefRef{SS15c1}{https://www.dropbox.com/sh/aads9hluqqzqfi6/AAAJ9qmq4Rj5B2LEbLsW88Esa/B12-1-s.mp4?dl=0}
\DefRef{SS15c2}{https://www.dropbox.com/sh/aads9hluqqzqfi6/AAD6AizEJ11HtFls21PjL3--a/B12-2-s.mp4?dl=0}
\DefRef{SS15c0}{https://www.dropbox.com/sh/aads9hluqqzqfi6/AADuH74Msrei2KFUhC4RsOGKa/B12-3-s.mp4?dl=0}
\DefRef{SS16c1}{https://www.dropbox.com/sh/aads9hluqqzqfi6/AADcQ5df0fo7tM8DBU1yLjK8a/B13-1-s.mp4?dl=0}
\DefRef{SS16c2}{https://www.dropbox.com/sh/aads9hluqqzqfi6/AACT2nQkH4U6tcMptvohtwhPa/B13-2-s.mp4?dl=0}
\DefRef{SS16c0}{https://www.dropbox.com/sh/aads9hluqqzqfi6/AAC1vTy8A7u5sNgtP3OIgmZka/B13-3-s.mp4?dl=0}

\usepackage{soul}
\usepackage{media9}
\usepackage{hyperref}
\graphicspath{{./}{figs/}{/scratch/vit/Data/BW/BW-img/}}
\DeclareGraphicsExtensions{.pdf,.png,.eps}
\usetikzlibrary{positioning}
\newlength{\fw}
\newlength{\ws}
\def\fbt{\fbox{\hspace*{0.975\ws}\vbox to 0.975\ws{}}}
\usepgfmodule{oo}
\pgfooclass{tmplt} {
  \method tmplt() { }
  \method apply(#1,#2,#3,#4,#5,#6) {
    \fill [black!0!white] (-2.55, 2.55) rectangle (2.55,-7.55);
    \node[inner sep=0pt,opacity=1.0] (v00) at (0,0)%
      {\includemedia[playbutton=none,width=\ws,flashvars={source=#1&autoPlay=true&loop=true}] {
        \includegraphics[angle=0,width=\ws,clip,trim=150 150 150
          150,decodearray={0 1 0 1 0 1 0 1}]{#2}}{VPlayer9.swf}}; 
    \node[black] at (0,0) {\href{#3}{\fbt}};
    \node[inner sep=0pt,opacity=1.0] (v01) at (0,-5)%
      {\includemedia[playbutton=none,width=\ws,flashvars={source=#4&autoPlay=true&loop=true}] {
        \includegraphics[angle=0,width=\ws,clip,trim=150 150 150
          150,decodearray={0 1 0 1 0 1 0 1}]{#5}}{VPlayer9.swf}}; 
    \node[black] at (0,-5) {\href{#6}{\fbt}};
    \node[inner sep=0pt,opacity=1.0] at (0,2.3) {\bf Complete wave trajectory};
    \node[inner sep=0pt,opacity=1.0] at (0,-2.7) {\bf Emergent loop pattern};
  }
}

\begin{document}

\title{Emergence of localized persistent weakly--evanescent cortical brain wave loops} 

\author{Vitaly L. Galinsky}
\email{vit@ucsd.edu}
\affiliation{Center for Scientific Computation in Imaging,
University of California at San Diego, La Jolla, CA 92037-0854, USA}
\affiliation{Department of ECE, University of California, San Diego,
  La Jolla, CA 92093-0407, USA}
\author{Lawrence R. Frank}
\email{lfrank@ucsd.edu}
\affiliation{Center for Scientific Computation in Imaging,
University of California at San Diego, La Jolla, CA 92037-0854, USA}
\affiliation{
Center for Functional MRI,
University of California at San Diego, La Jolla, CA 92037-0677, USA}

\date{\today}

\begin{abstract}
An inhomogeneous anisotropic physical model of the brain cortex is
presented that predicts the emergence of non--evanescent (weakly
damped) wave--like modes propagating in the thin cortex layers
transverse to both the mean neural fiber direction and to the cortex
spatial gradient.  Although the amplitude of these modes stays below
the typically observed axon spiking potential, the lifetime of these
modes may significantly exceed the spiking potential inverse decay
constant. Full brain numerical simulations based on parameters
extracted from diffusion and structural MRI confirm the existence and
extended duration of these wave modes.  Contrary to the commonly
agreed paradigm that the neural fibers determine the pathways for
signal propagation in the brain, the signal propagation due to the
cortex wave modes in the highly folded areas will exhibit no apparent
correlation with the fiber directions.  The results are
consistent with numerous recent experimental animal and human brain
studies demonstrating the existence electrostatic field activity in
the form of traveling waves (including studies where neuronal
connections were severed) and with wave loop induced peaks observed
in EEG spectra. The localization and persistence of these cortical
wave modes has significant implications in particular for neuroimaging
methods that detect electromagnetic physiological activity, such as
EEG and MEG, and for the understanding of brain activity in general,
including mechanisms of memory.
\end{abstract}

\maketitle

The majority of approaches to characterizing brain dynamical
behavior are based on the assumption that signal propagation along
well known anatomically defined pathways, such as major neural fiber
bundles, tracts or groups of axons
(down to a single axon connectivity) should be sufficient
to deduce the dynamical characteristics of brain activity at different
spatiotemporal scales.  Experimentally, data on which this assumption
is employed range from high temporal resolution neural oscillations
detected at low spatial resolution by EEG/MEG
\citep{pmid23367980,*pmid21167841} to high spatial resolution
functional MRI resting state modes oscillation detected at low
temporal resolution \citep{2014NatSR...4E6893L,*pmid24982140}. As a
consequence, a great deal of research activity is directed at
the construction of connectivity maps between different
brain regions (e.g., the Human Connectome Project
\citep{VanEssen:2013}), and using those maps to study dynamical
network properties with the help of different models of signal
communication through this network along structurally aligned pathways
\citep{Tuch:1999,*Haueisen:2002,*0031-9155-50-16-009}, i.e. by
allowing input from different scales or introducing axon propagation
delays \citep{pmid24505628}.

However, recent detection \citep{pmid27855061} of cortical wave
activity spatiotemporally organized into circular wave-like patterns
on the cortical surface, spanning the area not directly related to any
of the structurally aligned pathways, but nevertheless persistent over
hours of sleep with millisecond temporal precision, presents a
formidable challenge for network theories to explain such a remarkable
synchronization across a multitude of different local
networks. Additionally, many studies show evidence that
electrostatic field activity in animal or human hippocampus (as well
as cortex) are traveling waves
\citep{pmid19489117,*pmid29887341,*pmid29563572} that can
affect neuronal activity by modulating the firing rates
\citep{pmid3720881,*pmid1521610,*pmid21414915} and may possibly play an
important functional role in diverse brain structures
\citep{pmid20508749}. And perhaps more importantly it
has been experimentally shown
\citep{pmid26631463,*pmid24453330,*pmid30295923,*pmid30776338} that
periodic activity can self-propagate by endogenous electric fields
even through a physical cut \textit{in vitro} that destroys all
mechanisms of neuron to neuron communication.

In this paper we investigate a more general physical wave mechanism
that allows cortical surface wave propagation in the cross fiber
directions due to the interplay between tissue inhomogeneity
and anisotropy in the thin surface cortex layer.  This new mechanism
has been overlooked by previous models of brain wave characterization
and thus is absent from current network pathway reconstruction and
analysis approaches.

The main claim of this paper is that there is a simple and elegant
physical mechanism behind the existence of these cross-fiber waves
that can explain the emergence and persistence of wave loops and wave
propagation along the highly folded cortical regions with a relatively
slow damping.  The lifetime of these wave--like cortex activity events
can significantly exceed the decay time of the typical axon action
potential spikes and thus can provide ``memory--like'' response in the
cortical areas generated as a result of ``along--the--axon'' spiky
activations. This new mechanism may provide an alternative approach
for the integration of microscopic brain properties and for the
development of a ``physical'' model for memory.

The paper derives cortex wave dispersion relation from well known
relatively basic physical principles and provides illustrations of why
and how cortical tissue inhomogeneity and anisotropy influence
propagation magnitude, time-scales, and directions and supports
extended and highly structured regions of existence in dissipative
media using simple 1- and 2-D anisotropic models, as well as more
realistic full brain model based on a set of parameters extracted from
real diffusion and structural MRI acquisitions. The wave properties
(frequency ranges, phase and group velocities, possible spectra) are
then compared with real EEG wave acquisitions. Finally, examples of
wave propagation are studied analytically and numerically using a
simple idealized but informative spherical shell cortex model
(i.e.~thin inhomogeneous layer around a sphere with homogeneous
anisotropic conducting medium) as well as a more realistic anisotropic
inhomogeneous full brain model with actual cortical fold geometry that
clearly shows the emergence of localized persistent wave loops or
rotating wave patterns at various scales, including at 
similar scales of global rotation recently detected
experimentally \citep{pmid27855061}.

An important aspect of the cortical waves model we present
is that it is based on relatively simple but
physically motivated \textit{averaged} electrostatic properties of
human neuronal tissue within realistic data-derived brain tissue
distributions, geometries, and anisotropy.  While the source of
these averaged tissue properties includes the extraordinarily
complex network of neuronal fiber connections supporting a multitude
of underlying cellular, subcellular and extracellular processes, we
demonstrate that the inclusion of such details for the
activation/excitation process is not necessary to produce coherent,
stable, macroscopic cortical waves.  Instead, a simple and
elegant physical model for wave propagation in
a thin dissipative inhomogeneous and
anisotropic cortical layer of these averaged properties 
is sufficient to predict the emergence of
coherent, localized, and persistent wave loop patterns in the brain.

We will start with the most general form of description of brain
electromagnetic activity using Maxwell equations in a medium
\makeatletter%
\if@twocolumn%
  \begin{align}
  \nabla\cdot\mvec{D} &= \rho,\quad
  \nabla\times\mvec{H} = \mvec{J} + \D{\mvec{D}}{t}\quad \Rightarrow \quad
  \D{\rho}{t} &+ \mvec{\nabla}\cdot\mvec{J} = 0.\nonumber
  \end{align}
\else
  \begin{align}
  \nabla\cdot\mvec{D} &= \rho,\quad
  \nabla\times\mvec{H} = \mvec{J} + \D{\mvec{D}}{t}\quad \Rightarrow \quad
  \D{\rho}{t} + \mvec{\nabla}\cdot\mvec{J} = 0.\nonumber
  \end{align}
\fi
\makeatother

Using the electrostatic potential $\mvec{E}=-\nabla \phi$,
Ohm's law $\mvec{J}=\mvec{\sigma}\cdot\mvec{E}$ (where
$\mvec{\sigma}\equiv\{\sigma_{ij}\}$ is an anisotropic conductivity
tensor), a linear electrostatic property for brain tissue
$\mvec{D}=\varepsilon\mvec{E}$, assuming that the permittivity is a
``good'' function (i.e. it does not go to zero or infinity everywhere)
and taking the change of variables $\partial x \to \varepsilon
\partial x^\prime$, the charge continuity equation for the
spatial-temporal evolution of the potential $\phi$ can be written in
terms of a permittivity scaled conductivity tensor
$\bm{\Sigma}=\{\sigma_{ij}/\varepsilon\}$ as

\begin{align}
\label{eq:phiSigma}
\D{}{t} \left(\nabla^2 \phi \right) &=
-\mvec{\nabla}\cdot\mvec{\Sigma}\cdot \nabla\phi + \mathcal{F}, 
\end{align}
where we have included a possible external source (or forcing) term
$\mathcal{F}$.  For brain fiber tissues the conductivity tensor
$\mvec{\Sigma}$ might have significantly larger values along
the fiber direction than across them. The charge continuity
without forcing i.e.,~($\mathcal{F}=0$) can be written in
tensor notation as
\begin{align}
\label{eq:phiTensor}
\partial_t \partial_i^2 \phi &+ \Sigma_{ij}\partial_i\partial_j \phi +
\left(\partial_i\Sigma_{ij}\right)\left(\partial_j\phi\right)=0,
\end{align}
where repeating indices denotes summation. Simple linear wave
analysis, i.e.~substitution of $\phi\sim
\exp{(-i(\mvec{k}\cdot\mvec{r}-\Omega t)})$, gives the following
complex dispersion relation
\begin{align}
\label{eq:disp}
D(\Omega,\mvec{k}) = -i\Omega k_i^2 - \Sigma_{ij}k_i k_j - i \partial_i\Sigma_{ij} k_j = 0,
\end{align}
which is composed of the real and imaginary
components:
\begin{align}
\label{eq:gamma_omega}
\gamma \equiv \Im \Omega = \Sigma_{ij}\frac{k_i k_j}{k^2} \qquad
\omega \equiv \Re \Omega =-\frac{\partial_i\Sigma_{ij} k_j}{k^2}
\end{align}
The condition for non-- (or weak--) evanescence is that the
oscillatory (i.e., imaginary) component of $\phi$, characterized by
the frequency $\omega$, is much larger than the decaying (i.e.,
real) component, characterized by the damping $\gamma$: i.e. that
the condition $|\gamma/\omega| \ll 1$ must be satisfied. This
requirement is clearly not satisfied if reported average isotropic and
homogeneous parameters are used to describe brain tissues.  For
typical low frequency ($\lesssim 10Hz$) white and gray matter
conductivity and permittivity (i.e. from 
\citep{pmid8938025,*pmid8938026}) $\varepsilon_{GM}=4.07\cdot
10^{7}\varepsilon_0$, $\varepsilon_{WM}=2.76\cdot
10^{7}\varepsilon_0$, $\sigma_{GM}=2.75\cdot 10^{-2}$ S/m,
$\sigma_{WM}=2.77\cdot 10^{-2}$ S/m, where
$\varepsilon_0=8.854187817\cdot 10^{-12}$ F/m is the vacuum
permittivity) the damping rate $\gamma$ is in the range of
$75$--$115$ s\textsuperscript{-1} which would be expected to give
strong wave damping. 

\begin{figure}[!tbh] \centering
\includegraphics[width=1.0\columnwidth]{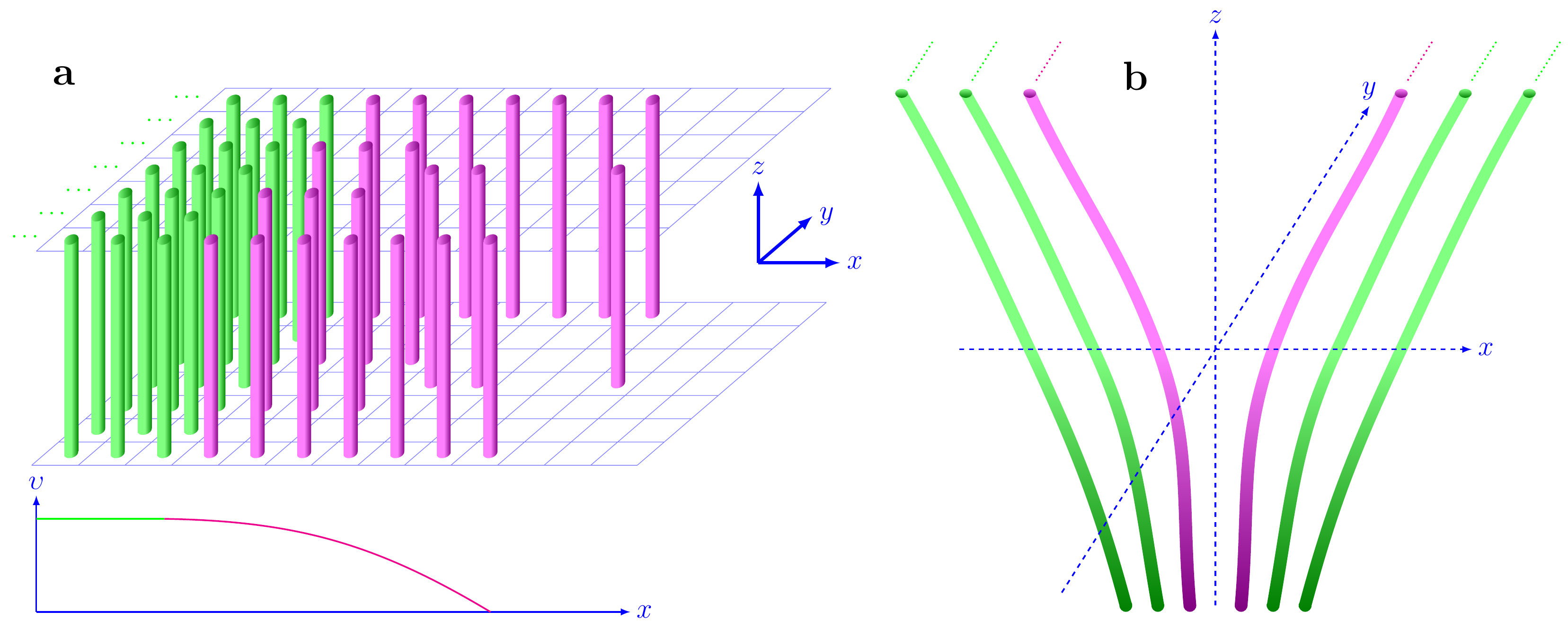}
\caption[]{(a) Schematic picture of half-plane packing of fibers. The
  uniform area of fibers oriented along $z$ direction (shown in green)
  is bounded by a thin transitional area (magenta) where the
  conductivity gradient may be important (a sketch of one
  possible conductivity profile is shown in the bottom panel).  (b)
  Schematic picture that can be used as a crude two dimensional
  approximation of fold. The direction of fiber conductivity has only
  $x$ and $z$ components and all quantities are assumed to be uniform
  in $y$ direction.  }
\label{fig:fibers}
\end{figure}

In order to better understand the effects of brain tissue
micro- and macro-structure on the manifestation of propagating brain
waves, it is instructive to consider two idealized tissue models.
In the first model (\cref{fig:fibers}a)
all brain fibers are packed in a
half space aligned in $z$ direction and their number decreases in $x$
direction in a relatively thin layer at the boundary.
We assume
that small cross fiber
currents can be characterized by a small parameter $\epsilon$ and
introduce the conductivity tensor as
\begin{align}
\mvec{\Sigma}&=
\begin{pmatrix}
\epsilon \upsilon & \epsilon \upsilon & \epsilon \upsilon\\
\epsilon \upsilon & \epsilon \upsilon & 0\\
\epsilon \upsilon & 0 & \upsilon
\end{pmatrix}.
\end{align}
where $\upsilon \equiv \upsilon(x)$. For the $\upsilon(x)$
dependence we will assume that the conductivity is changing only
through a relatively narrow layer at the boundary (as illustrated in
\cref{fig:fibers}a) and the conductivity gradient is directed along
$x$ axis. 
Then we will look for a solution for the potential $\phi$ located in
the thin layer of inhomogeneity (that is we substitute $x\to\epsilon
x$) that depends on $z$ and $y$ only as $\phi = \phi_{\parallel}(z) + \epsilon
\phi_{\perp}(y)$.

\begin{align}
\epsilon^0: &\qquad \D{}{t} \D{^2\phi_{\parallel}}{z^2} +\upsilon
\D{^2\phi_{\parallel}}{z^2} + \D{\upsilon}{x}\D{\phi_{\parallel}}{z} = 0
\label{eq:eqZ}
\\
\epsilon^1: &\qquad \D{}{t} \D{^2\phi_{\perp}}{y^2} +
\D{\upsilon}{x}\D{\phi_{\perp}}{y} + O(\epsilon) = 0
\label{eq:eqY}
\end{align}
where $\epsilon^0$ and $\epsilon^1$ denote the zeroth
and the first orders of $\epsilon$ power.

The first equation \cref{eq:eqZ} describes a potential along the fiber
direction and is a damped oscillator equation that has
a decaying solution.
But the second
equation \cref{eq:eqY} describes a potential perpendicular to the
fiber direction and does not include a damping term, hence it
describes a pure wave-like solution 
that propagates in the thin layer transverse to the main fiber
direction. Thus although this wave-like solution
$\phi_{\perp}$  has a smaller amplitude
than along the fiber action potential $\phi_{\parallel}$, it
can nevertheless have a much longer lifetime.

We would like to stress that the equations \cref{eq:eqZ,eq:eqY} are 
here only for illustrative purposes to reiterate a relatively obvious 
but often overlooked consideration that under anisotropic inhomogeneous 
conditions some direction may happen to be better suited for wave propagation
than the others. Those equations should not be considered as
the most important part as the complete dispersion relation \cref{eq:disp}
will be used to study waves propagation later.

In order to account for geometric variations, we construct a slightly
more complex two dimensional model that can be viewed as a very crude
approximation of a cortical fold (\cref{fig:fibers}b).
The equation for the
$\phi_{\perp}(y)$ will again be of a wave type, similar to
\cref{eq:eqY}, with the addition of a
$z$ component of the conductivity gradient and with a similar
wave--like solution that will include 
an additional term induced by the inhomogeneity in $z$.

The sole purpose of those examples is to provide illustrations that
dissipative media with complex structure may show 
surface wave--like solutions. Surface waves at
the boundary of various elastic media have been
extensively studied and used in various areas of science (including
acoustics, hydrodynamics, plasma physics, etc.) since the work of Lord
Rayleigh
\citep{citeulike:3581102}.  The
existence of surface waves at the dissipative medium boundary is also
known \citep{2014PlPhR..40..650K}.

In order to extend the above analysis to a more realistic model of
brain tissue architecture we extracted volumetric structural brain
parameters from high resolution anatomical MRI datasets as well as
brain fiber anisotropy from diffusion weighted MRI datasets. All
anatomical and diffusion MRI datasets are from the Human Connectome
Project \citep{VanEssen:2013}.  The details for all of the
processing steps can be found in
\citep{Galinsky:2014,Galinsky:2015,pmid30230014}.  More refined
procedures for constructing the conductivity tensor and anisotropy in
different brain regions, cortical areas in particular, would clearly
be beneficial and will be addressed in the future.

\begin{figure}[!tbh] \centering
\includegraphics[width=0.9\columnwidth]{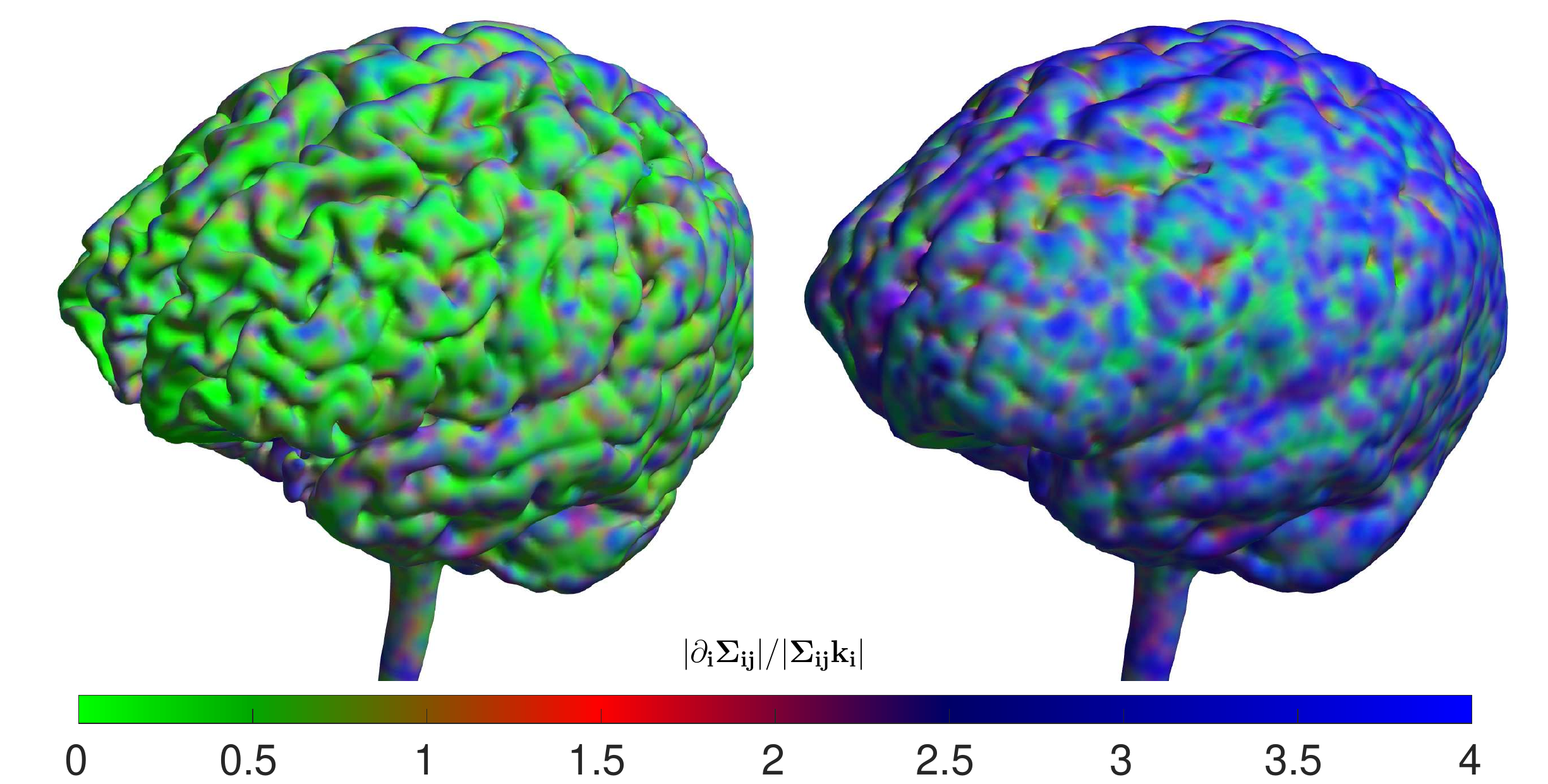}
\caption[]{Isovolume maps for comparison of dissipation vs wave--like
  effects at different cortex layers. The left panel shows the
  isosurface from the cortex area that may be representative of
  white--gray matter interface. The right panel shows the isosurface
  that is located in the outer cortex area of gray matter. The color
  scheme uses shades of green to mark regions where dissipative term
  dominates, i.e.~$|\Sigma_{ij}k_i|\ge|\partial_i\Sigma_{ij}|$, shades
  of red where $|\Sigma_{ij}k_i| < |\partial_i\Sigma_{ij}|\le
  2|\Sigma_{ij}k_i|$, and shades of blue where the wave-like term is
  more than two times dominant,
  i.e.~$2|\Sigma_{ij}k_i|<|\partial_i\Sigma_{ij}|$. The inner cortex
  shown in the left panel clearly display prevalence of dissipation,
  whereas the outer cortex shown in the right panel allows for
  wave--like cortex activity in a majority of the locations.  }
\label{fig:ratio}
\end{figure}

To provide an illustration of where the conductivity anisotropy and
inhomogeneity can form appropriate conditions for cortex surface waves
generation we created plots that can be used to characterize the ratio
$|\gamma/\omega|$. To construct the ratio we calculated two vectors,
$\partial_i\Sigma_{ij}$ and $\Sigma_{ij}k_i$ and compared their
norms. For wave vector $\mvec{k}$ we used a vector with the same
direction as in $\partial_i\Sigma_{ij}$ vector and magnitude
$|\mvec{k}|=|\nabla\mvec{\Sigma}|/|\mvec{\Sigma}|$, i.e.~our intention
is to compare norms of dissipative and wave-like terms at $kh\approx
1$ where $h$ is the cortical thickness. \Cref{fig:ratio}
shows plot of the ratio of $|\Sigma_{ij}k_i|$ and
$|\partial_i\Sigma_{ij}|$ at two different depths inside the brain
with dissipative regime in the inner cortex (left) and wave--like
conditions in the outer cortex (right).

Both anisotropy and inhomogeneity are important for the existence of
the cortex surface waves. For example for inhomogeneous but isotropic
tissue the conductivity tensor $\Sigma_{ij}$ will simply be $S
\delta_{ij}$, where $S$ is a scalar inhomogeneous
conductivity. Therefore both the phase velocity $(v_{ph} =
\omega/k)$ and the group velocity $(v_{gr} =
\partial \omega/\partial k)$ will include terms $\nabla S$ and
$\nabla S\cdot\mvec{k}$, meaning that those waves are not able to
propagate normally to the local conductivity gradient $\nabla S$. This
restriction is absent in cortex areas when both anisotropy and
inhomogeneity are present.

To provide some (possibly overly optimistic) estimates based on a
typical human brain dimensions we can consider a simple spherical
cortex shell model with a cortical layer of fixed thickness $h\approx$
1.5--3mm spread over a hemisphere of radius $R\approx$ 75mm, with all
parameters kept constant inside the hemisphere (for $r<R$) and
changing as a function of radius $r$ in a cortical layer. Even without
taking into account the known strong anisotropy of
neural tissue, these simple geometric
considerations provides for the longest waves (with the
smallest amount of damping) with $|\gamma/\omega| \sim k h
\sim$ 0.02--0.04.  Anisotropy will reduce this estimate
even further, thus further strengthening the condition
necessary to support stable waves. This simple spherical shell
model can be used to illustrate the natural
appearance of standing--type cortical waves, which
can be easily understood from simple geometrical optics
arguments. Using the dispersion relation \cref{eq:disp} with only
single component of the conductivity tensor $\Sigma_{zz} \equiv S(r)$,
the wave frequency $\omega$ can be expressed as
\begin{align}
\label{eq:freq}
\omega = -\frac{1}{r}\dd{S}{r} \frac{k_z z}{k^2};
\end{align}
where we have neglected the term $Sk_z^2$ because $kh\ll
1$. Then from the geometrical optics ray equations
\begin{align}
\label{eq:rays}
\dd{r_l}{t} = \D{\omega}{k_l},\qquad
\dd{k_l}{t} =-\D{\omega}{r_l},
\end{align}
we can get
\begin{equation}
\label{eq:simple_loop}
\def\arraystretch{2.2}
\begin{array}[c]{*4{>{\displaystyle}l}}
\dd{x}{t} &=-2\omega\frac{k_x}{k^2},\qquad
&\dd{k_x}{t} &=-\dd{\omega}{r}\frac{x}{r},\\
\dd{y}{t} &=-2\omega\frac{k_y}{k^2},\qquad
&\dd{k_y}{t} &=-\dd{\omega}{r}\frac{y}{r}, \\
\dd{z}{t} &=-\omega\frac{2 k_z^2-k^2}{k^2 k_z},\qquad
&\dd{k_z}{t} &=-\frac{1}{z}\left[\dd{\omega}{r}\frac{z^2}{r}+\omega\right].
\end{array}
\end{equation}
These equations will generate rays inside the spherical cortex shell
showing wave propagation across both the fibers and the conductivity
gradient in the cortex subregion where $\omega \, d\omega/dr < 0$. For
$k_z=k/\sqrt{2}$ and $z=\sqrt{-\omega r/(d\omega/d r)}$ the wave path
has the simplest form -- the wave follows the same loop through the
cortex over and over again.  Different families
of frequencies $\omega$ and wavevectors $\mvec{k}$ will result in the
appearance of cortical wave loops at different cortex locations.

\begin{figure}[!tbh] \centering
\includegraphics[width=0.95\columnwidth]{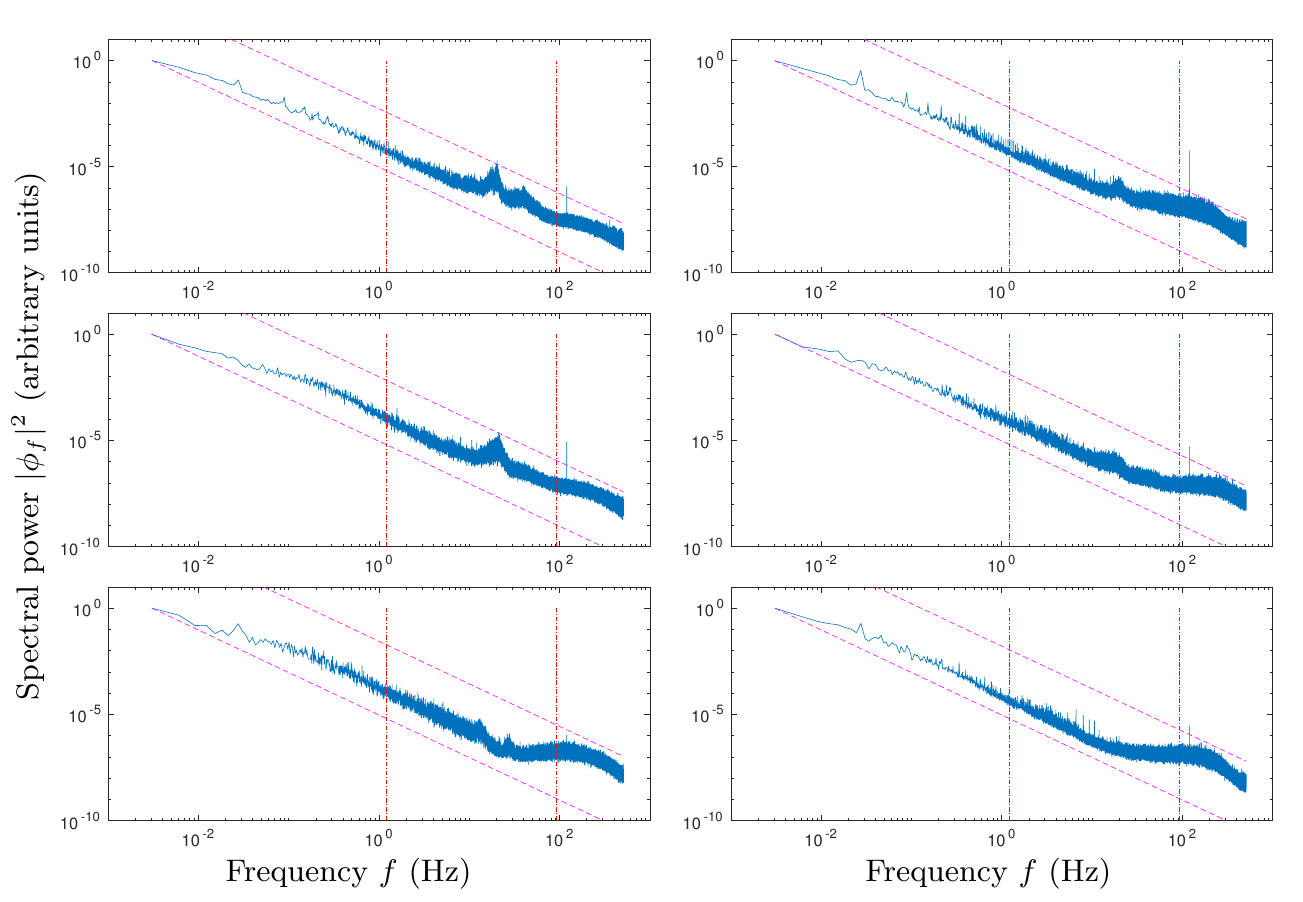}
\caption[]{Spectral power of EEG signal collected with 64 sensor array
  and averaged over all sensors for six independent subjects is shown
  in six panels. The dashed lines outline the predicted $f^{-2}$ in
  the lower ($f\lesssim 1.2$Hz) and higher ($f\gtrsim 92$Hz) parts of
  the spectra. The dashed--dotted vertical lines denote the
  frequency range where the cortical wave loops may be generated. Both
  the slope and the range agree very well with typically observed
  values.
\label{fig:ps}}
\end{figure}

In order to determine a possible energy distribution across different
frequencies for these cortical waves some knowledge about the forcing
term $\mathcal{F}$ in \cref{eq:phiSigma} is required. A rough estimate
for this distribution in the form of a power spectra scaling can be
carried out using some simple assumptions. Assuming that the forcing
consists of spiking input localized at random locations and times, it
can be described as a sum of delta functions, $\mathcal{F}=\sum_i A_i
\delta(t-t_i)\delta(\mvec{r}-\mvec{r}_i)$, corresponding to a flat
forcing frequency spectrum in the Fourier domain.  Then, from the last
term in \cref{eq:phiTensor} $|(\partial_i\Sigma_{ij})(\partial_j
\phi)| \sim \omega \sqrt{|\phi_\omega|^2} \sim \mathrm{const}$, we can
estimate the exponent $\alpha=2$ in a power law scaling of the
potential $\phi$ frequency spectrum (i.e.~for $|\phi_\omega|^2 \sim
\omega^{-\alpha}$).

The presence of the cortical wave loops described above can modify
this $\omega^{-2}$ dependence and thus modify the spectrum in such a
way that spectrum peaks are generated that correspond to these loop
wave currents.  From \cref{eq:freq} we can estimate the range of
frequencies where those loops can possibly be present. Taking $1/r\sim
1/R$, $dS/dr \sim \Sigma_{zz}/h$, $z=\sqrt{-\omega
  r/(d\omega/dr)}\sim\sqrt{Rh}$ gives a frequency estimate as
$f=\omega/2\pi \sim \Sigma_{zz}/(2\pi k\sqrt{2R h})$, hence for the
largest and smallest wavenumbers defined by the smallest (1.5mm) and
largest (75mm) loop radii, the frequency $f$ spans the range
$1.2$--$92\mathrm{Hz}\ (v_{ph}\approx0.002-7\mathrm{m/s})$. The large
scale circular cortical waves of \citep{pmid27855061} ($9$--$18$Hz,
$2$--$5$m/s) are clearly inside this range.  The above spherical shell
wave propagation example (as well as wave simulations presented below)
are not able to predict the exact values for wave amplitude as it
requires calculation of the balance between excitation and dissipation
of these waves at various frequency ranges and the simple estimates
using amplitudes of spiking are not that 
particularly informative as they simply
predict the values that are in the range of typically observed field
potentials. Nevertheless, even without the amplitude estimation our
analysis is able to predict the preferred directions of wave
propagation and loop pattern formation based on geometrical and tissue
properties.  Moreover, this framework provides the mechanisms
to incorporate a more complete quantitative description of axonal
spiking, which is beyond the scope of this paper but currently under
investigation in our lab.

Evidence for the existence of these wave loop induced spectral peaks
is shown in \Cref{fig:ps}, which shows the spectral power of the EEG
signal for six subjects \cite{pmid26190988,*pmid12433379}, averaged
over all sensors.  The dashed lines outline the predicted $f^{-2}$ in
the lower ($f\lesssim 1.2$Hz) and higher ($f\gtrsim 92$Hz) parts of
the spectra. The dashed--dotted vertical lines denote the
frequency range where the cortical loops may exist, agreeing very well
with typically observed EEG excessive activity range, from
low frequency delta (0.5--4Hz) to high frequency gamma (25--100Hz)
bands.

\begin{figure}[!tbh] \centering
\def\fsc{0.55}
\setlength{\ws}{0.15\textwidth}
\includegraphics[width=0.99\columnwidth]{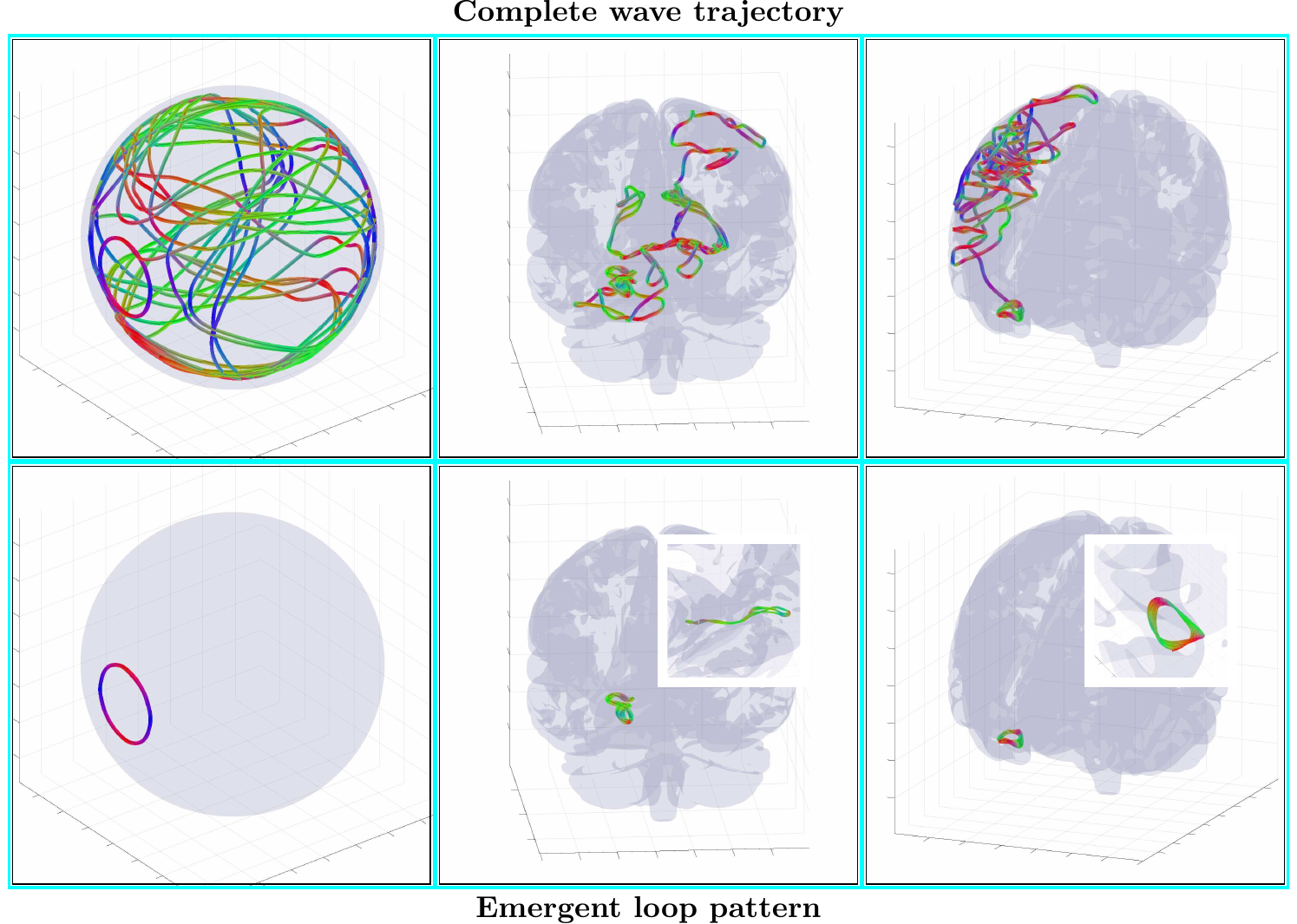}
\caption[]{ Complete wave packet trajectory snapshots (upper row) and
  emergent stable wave loop patterns (lower row) for the thin
  spherical shell cortex model (left) and for the realistic cortex
  fold geometry (middle and right). All
  wave packets were initialized with random parameters assuming the
  presence of spiky activation sources (not shown). 
  Movie files for these and
  additional loop examples can be found in \cite{FigShare}.
\label{fig:rays}}
\end{figure}

The effect described theoretically above can be demonstrated through
numerical simulation of wave propagation in a thin dissipative
inhomogeneous and anisotropic cortical layer.  As a starting point for
numerical study we included \cref{fig:rays}, that shows spatial
snapshots of the dynamical behavior of randomly generated
wave trajectories (the movies are in \cite{FigShare}) in a
regime with $\gamma_{ef\!f} L_{loop}/v_{gr} \le 1$ (where
$\gamma_{ef\!f}$ is an effective wave dissipation rate, i.e. a
difference between average dissipation and spiky activations rates for
a wave packet propagating with group velocity $v_{gr}$ along some
characteristic loop of $L_{loop}$ length.  Wave packets are simulated
using the ray equations \cref{eq:rays} where the general form of
anisotropic dispersion relation \cref{eq:disp,eq:gamma_omega} is used.

In the idealized spherical model some of the emergent persistent
localized cortical wave patterns are precisely the simple loop pattern
predicted by \cref{eq:simple_loop}, despite the complex initial
spatiotemporal pattern of the initial wave trajectory.  As further
predicted, the simulations informed by the real human data 
with the same cortex fold geometry
as in \cref{fig:ratio} (middle
and right panels) also produce stable loop structures, now embedded
within the complex geometry of the cortical folds.

All wave simulations were initialized with wave packets of random
parameters (frequency, wave number, location, etc), but clearly show
an emergence of localized persistent closed loop patterns at different
spatial and temporal scales, from 
scales as large as global the whole brain
rotational wave activity experimentally detected in
\citep{pmid27855061} to as
small as the resolution used for the cortical layer thickness
detection.

In conclusion, in this paper we have presented an inhomogeneous
anisotropic physical model of wave propagation in the brain
cortex. The model predicts that in addition to the well-known damped
oscillator--like wave activity in brain fibers, there is another class
of brain waves that are not directly related to major fibers, but
instead propagate perpendicular to the fibers along the highly folded
cortical regions in a weakly--evanescent manner that results in their
persistence on time scales long compared to waves along brain fibers.
Thus waves can potentially propagate in any direction,
including the direction along the fibers or in the direction of the
inhomogeneity gradient. However, the dissipation of the waves is
smallest when they propagate cross fibers, therefore, on average the
cross fiber direction of propagation should be seen more often. For
the first time, we have obtained the dispersion relation for those
surface cortex waves, and have shown, both analytically and
numerically, a plausible argument for their existence. Through
numerical analysis we have developed a procedure to generate an
inhomogeneous and anisotropic distribution of conductivity tensors
using anatomical and diffusion brain MRI data. While the detailed
numerical studies of effects and importance of these waves and their
possible biological role are out of the scope of this paper, we have
presented preliminary results that suggest that the time of life for
these wave--like cortex activity events may significantly exceed the
decay time of the typical axon action potential spikes. Thus, they can
provide an persistent neuronal response in the cortical areas
generated as a result of ``along--the--axon'' spiky activations.

The interaction of the traveling waves with spiking activity
is of course an interesting and important question.  Our model
supports the inclusion of spiking activity as a source term to the
wave model (one example is sketched in \cref{sec:spiking}.
In particular this may allow the
development of models of brain rhythm generation based on coupling
with spiking sources.  Exploration of the interplay of the traveling
waves and spiking activity with this model will certainly be the
focus of future work.

The ranges of parameters for the waves produced by our model
are in agreement with presented by several studies
\citep{pmid19489117,*pmid29887341,*pmid29563572} that present
evidence that electrostatic field activity in several
areas of animal and human brains are traveling waves that can affect
neuronal activity by modulating the firing rates
\citep{pmid3720881,*pmid1521610,*pmid21414915} and may possibly play an
important functional role in diverse brain structures
\citep{pmid20508749}.  The natural self organization of these
traveling waves into loop--like structures that our model
produces agrees well with recently detected \citep{pmid27855061}
cortical wave activity spatiotemporally organized into circular
wave-like patterns on the cortical surface. Self-propagation of
endogenous electric fields through a physical cut in vitro when all
mechanisms of neuron to neuron communication has been destroyed
\citep{pmid26631463,*pmid24453330,*pmid30295923,*pmid30776338} can
potentially be attributed to these waves as well.  We have
also demonstrated that the peaks these wave loops would induce in
EEG spectra are consistent with those typically observed EEG data.

Direct experimental results have shown that even despite
the small amplitudes of the external field potentials
relative to the threshold of a spiking neuron, external
fields can play a substantial role in the spiking activity
and ``even very small and slowly changing fields that triggered $V_e$
changes under 0.2 mV led to phase locking of spikes to the external
field and to a greatly enhanced spike-field synchrony''
\citep{pmid21240273}.  Therefore our models ability
to predict regions of cortex where external wave
activity can emerge and form a sustained loop pattern
has the potential to be important for
understanding where the neuron spiking synchronization will have
better chances to be achieved, hence it can provide substantial input
in understanding effects on neural information processing and
plasticity.

The ability of this new physical model for the generation,
propagation, and maintenance of brain waves may have significant
implications for the analysis of electrophysiological brain recording
and for current theories about human brain function.  Furthermore, the
dependence of these waves on brain geometry, such as cortical
thickness, has potentially significant implications for understanding
brain function in abnormal states, such as Alzheimer's Disease, where
cortical thickness changes are evident, and the dependence on tissue
status may be important in conditions such as Traumatic Brain Injury,
where tissue damage may alter its anisotropic and inhomogeneous
properties.

\section*{Acknowledgements}
LRF and VLG were supported by NSF grants DBI-1143389,
DBI-1147260, EF-0850369, PHY-1201238, ACI-1440412, ACI-1550405 and
NIH grant R01 MH096100.  Data were provided [in part] by the Human
Connectome Project, WU-Minn Consortium (Principal Investigators:
David Van Essen and Kamil Ugurbil; 1U54MH091657) funded by the 16
NIH Institutes and Centers that support the NIH Blueprint for
Neuroscience Research; and by the McDonnell Center for Systems
Neuroscience at Washington University. Data collection and sharing
for this project was provided by the Human Connectome Project (HCP;
Principal Investigators: Bruce Rosen, M.D., Ph.D., Arthur W. Toga,
Ph.D., Van J. Weeden, MD). HCP funding was provided by the National
Institute of Dental and Craniofacial Research (NIDCR), the National
Institute of Mental Health (NIMH), and the National Institute of
Neurological Disorders and Stroke (NINDS). HCP data are disseminated
by the Laboratory of Neuro Imaging at the University of Southern
California.  The EEG data is courtesy of Antigona Martinez of the
Nathan S.~Kline Institute for Psychiatric Research.

\onecolumngrid

\appendix

\renewcommand{\thefigure}{A\arabic{figure}}
\setcounter{figure}{0}
\def\fsc{1.6}

\section{Spherical shell cortex model}

We first provide details about
the procedures used for generating inhomogeneous and anisotropic
components of the permittivity scaled conductivity tensor
$\mvec{\Sigma}$. 

Spherical shell cortex model is represented by fixed anisotropy tensor
$\sigma_{ij}(\mvec{r})\equiv \sigma_{ij}$ scaled by a radial
inhomogeneous density $\rho(\mvec{r})\equiv \rho(r)$, such that the
total conductivity tensor $\Sigma_{ij}(\mvec{r})$ is defined by

\begin{equation}
\Sigma_{ij}(\mvec{r}) = \rho(\mvec{r}) \sigma_{ij}(\mvec{r}),
\label{eq::Sigma}
\end{equation}
where $\sigma_{ij}$ is a constant,  anisotropic, positive semidefinite
symmetric tensor and $\rho(r)$ is piecewise continuous function of
radius $r$ ($0\le r  \le 1$) 
\begin{align}
\begin{tikzpicture}
\node at (-5,0) {
\begin{minipage}{0.5\textwidth}
\begin{equation}
\rho(r) =
\begin{cases}
1 & r \le r_0 \\ 
\displaystyle\frac{1
+\arctan\left(\alpha_\infty \left(1 -
2\displaystyle\frac{r-r_0}{r_1-r_0}\right)\right)^{2n+1}}{
2\arctan\left(\alpha_\infty\right)^{2n+1}}
& r_0<r<r_1 \\ 
0 & r \ge r_1
\end{cases}\nonumber
\end{equation}
\end{minipage}
};
\node at (5,0) {
\includegraphics[width=0.4\columnwidth]{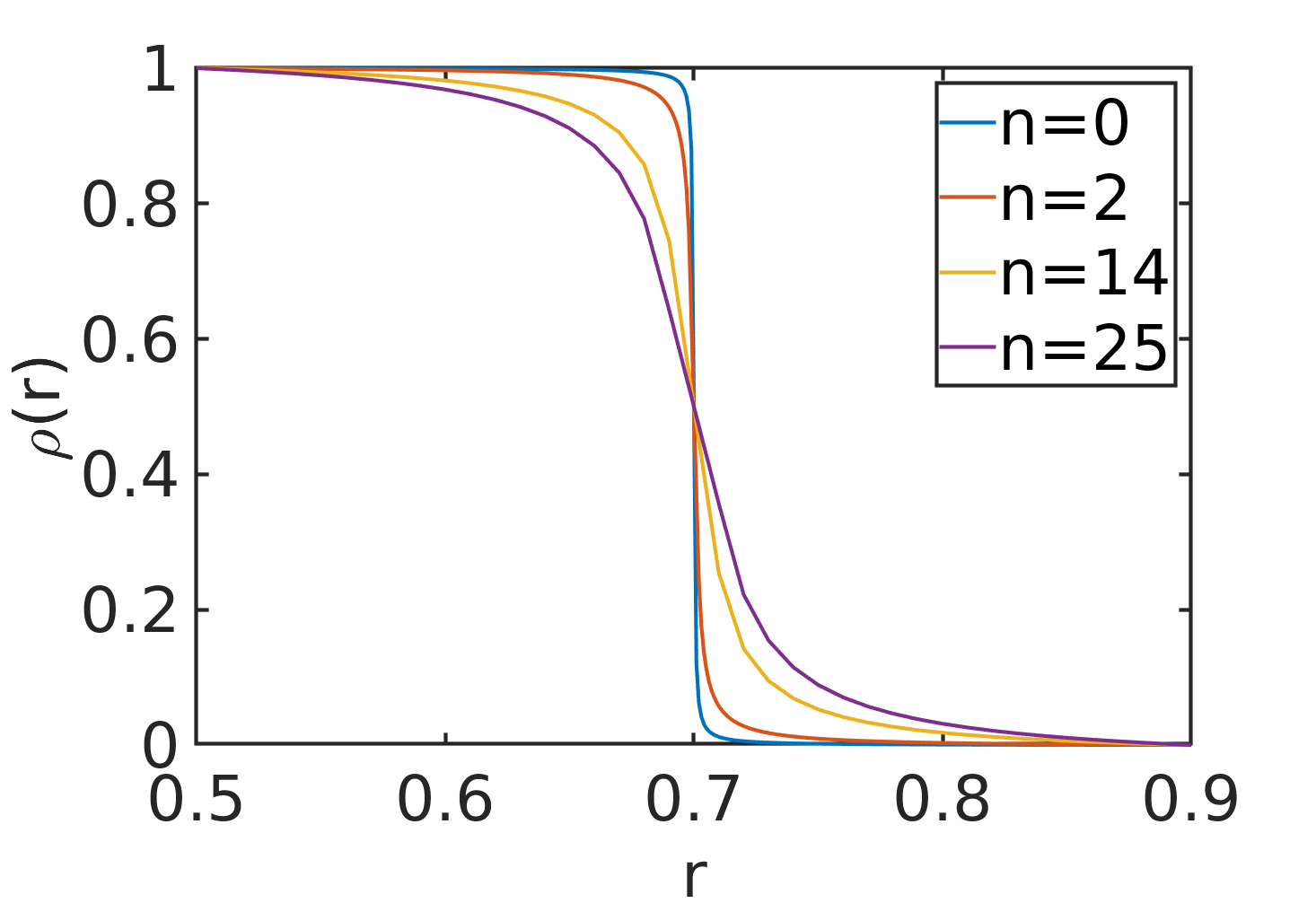}
};
\end{tikzpicture}
\end{align}

The spherical shell cortex wave simulations include three different
anisotropic tensor $\sigma_{ij}$ choices, 
\begin{align}
\mvec{\sigma}^{(1)}&=
\begin{pmatrix}
\varepsilon & 0           & 0\\
0           & \varepsilon & 0\\
0           & 0           & 1
\end{pmatrix} 
\qquad
\mvec{\sigma}^{(2)}=\frac{1}{3}
\begin{pmatrix}
1+2\varepsilon & 1-\varepsilon  & 1-\varepsilon\\
1-\varepsilon  & 1+2\varepsilon & 1-\varepsilon\\
1-\varepsilon  & 1-\varepsilon  & 1+2\varepsilon
\end{pmatrix} 
\qquad
\mvec{\sigma}^{(3)}=
\begin{pmatrix}
1  & 0           & 0\\
0  & \varepsilon & 0\\
0  & 0           & 1
\end{pmatrix} 
\label{eq::fixed_sigma}
\end{align}
where (for $\varepsilon = 0$) $\mvec{\sigma}^{(1)}$ represents the
conductivity tensor used in analytical solution of equations (10) of
the paper, i.e.~currents only in $z$ direction, $\mvec{\sigma}^{(2)}$
represents different orientation, where currents are allowed in the
direction of 45$^\circ$ relative to all axis, and
$\mvec{\sigma}^{(3)}$ represents more complicated current anisotropy,
with crossing currents flowing in $x$ and $z$ direction, but with no
currents in $y$ direction. 

Parameters $n$, $r_0$ and $r_1$ used to control the thickness of the
inhomogeneous ``cortical'' layer ($r_0$=0.5, $r_1$=0.9,
$\alpha_\infty$=500 and $n=$0,2,14
and 25 used for \crefrange{S1}{S4}). 

\section{Cortical fold model}

For cortical fold model the conductivity tensor
$\Sigma_{ij}(\mvec{r})$ is again defined as a product of inhomogeneous
$\rho(\mvec{r})$ and anisotropic $\sigma_{ij}(\mvec{r})$ parts
\cref{eq::Sigma}, but both parts are now functions of brain
locations. 

\subsection{Inhomogeneity estimation}

The inhomogeneous density function $\rho(\mvec{r})$ is estimated from
high resolution anatomical MRI data by processing it with SWD
\cite{Galinsky:2014} (skull stripping, field of view normalization, noise
filtering) and then registering to MNI152 space
\cite{pmid20656036,FONOV2009S102} with SYM-REG
\cite{pmid30230014}. The final 1mm$^3$ 182x218x182 volume is used as
the inhomogeneous density $\rho(\mvec{r})$. 

\subsection{Anisotropy estimation}

For the anisotropy tensor $\sigma_{ij}(\mvec{r})$ several different test cases were employed.

\subsubsection{Fixed anisotropy orientation and value} 

The same $\sigma^{(1)},\sigma^{(2)}$ and $\sigma^{(3)}$ fixed
(location independent) anisotropic tensors as in the spherical shell
cortex model \cref{eq::fixed_sigma} were assigned to every location in
the brain. This is the simplest case that may allow to separate the
effects of inhomogeneity and anisotropy on loop pattern formation. 

\subsubsection{Varying anisotropy orientation and fixed anisotropy value}

Multiple diffusion direction and diffusion gradient strength MRI data
were used to estimate diffusion tensor $D_{ij}$ \cite{pmid17946125}.  

The anisotropy orientation was estimated using eigenvector
$\mvec{d}^{(1)}$ of the diffusion tensor $D_{ij}$ with the largest
eigenvalue $\lambda_d^{(1)}$. The anisotropy tensor
$\sigma_{ij}(\mvec{r})$ was defined as  
\begin{equation}
\mvec{\sigma}(\mvec{r}) = \mvec{R}^T \mvec{\sigma}^{(1)} \mvec{R}
\label{eq::sigma_orientation}
\end{equation}
where the value of anisotropy is constant across the volume
($\varepsilon$ = 0, 0.01 and 0.1 were used for different test
examples).  $\mvec{R}$ is a rotation matrix between directions (0,0,1)
and $\mvec{d}^{(1)}\equiv (x, y, z)$ that can be expressed as
\begin{align}
\displaystyle
\mvec{R}&=
\begin{pmatrix}
1-\frac{x^2}{1+z} & -\frac{x y}{1+z}  & x\\
-\frac{x y}{1+z}  & 1-\frac{y^2}{1+z} & y\\
-x                & -y                & 1-\frac{x^2+y^2}{1+z}
\end{pmatrix} 
\end{align}

\subsubsection{Varying anisotropy orientation and value}

Similarly to the previous case the diffusion tensor $D_{ij}$ was
estimated for each voxel and eigenvector $\mvec{d}^{(1)}$ with the
largest eigenvalue $\lambda_d^{(1)}$ was used to define the anisotropy
tensor major axis. The position dependent anisotropy value was defined
by intoducing parameter $\varepsilon$ as either
$\lambda_d^{(2)}/\lambda_d^{(1)}$ or
$(\lambda_d^{(2)}+\lambda_d^{(3)})/2/\lambda_d^{(1)}$ (in both cases
resulting in axisymmetric form of the conductivity tensor) and also
using two different values $\lambda_d^{(2)}/\lambda_d^{(1)}$ and
$\lambda_d^{(3)}/\lambda_d^{(1)}$ for $\sigma_{22}$ and $\sigma_{11}$.

\subsubsection{Varying anisotropy orientation and value estimated from
  microstructure fiber anisotropy} 

The microstructure anisotropy at the level of a single cell and its
extracellular vicinity may be significantly higher than detected by
diffusion MRI estimates. Recent measurements of effective impedance
that involve both intercellular and extracellular electrodes
\cite{pmid26745426} show significantly lower conductivity (by orders
of magnitude) than conductivity obtained by measurements between
extracellular electrodes only. Although attributing this to lower than
typically assumed conductivity (or higher impedance) of extracellular
medium may be questionable \cite{pmid28978453}, this clearly confirms
high anisotropy for the effective conductivity that should be used for
analysis of effects of intercellular and membrane sources (and these
are the most important type of sources for generation of surface brain
waves considered in this paper). This may be important for future
generation and refinement of the macroscopic conductivity estimates
from microstructure data.

We assumed here that anisotropic form of $\sigma^{(1)}_{ij}$ with
$\varepsilon=0.001,0.01$ or 0.1 can be used to describe different
levels of intercellular conductivity as well as microstructure
extracellular conductivity in the vicinity of cell membranes (where
$z$-direction corresponds to the direction of the fiber). Using full
brain tractography results \cite{brain-grid} we generated the
anisotropic part of the conductivity tensor
$\mvec{\sigma}(\mvec{r})$ in voxel at $\mvec{r}$ location as an
average over all fiber orientations assuming $N$ fibers inside the
voxel with $\mvec{R}_k$ orientation matrix for every fiber $k$, i.e.
\begin{equation}
\mvec{\sigma}(\mvec{r}) = \frac{1}{N}\sum^N_k \mvec{R}^T_k
\mvec{\sigma}^{(1)} \mvec{R}_k,
\label{eq::sigma_sum}
\end{equation}
with the complete form of the conductivity tensor
$\mvec{\Sigma}$ again given by \cref{eq::Sigma}.

\section{Wave trajectory integration}

To obtain the trajectory of a brain wave with
frequency $\omega$ emitted at a point $\mvec{r}_0$ with an initial
wave vector $\mvec{k}_0$ we substitute the inhomogeneous anisotropic
conductivity tensor $\mvec{\Sigma}$ in the wave dispersion relation
(imaginary part of eq. (3) of the paper)
\begin{align}
\label{eq::dispI}
\mathcal{D}(\omega,\mvec{k}) = \omega k^2 +\partial_i\Sigma_{ij} k_j = 0,
\end{align}
and obtain wave ray equations as
\begin{align}
\label{eq::ray}
\dd{t}{\tau} &= \D{\mathcal{D}(\omega,\mvec{k})}{\omega} = k^2,\qquad
\Rightarrow\qquad t = \int_0 k^2 d\tau\\ 
\dd{\omega}{\tau} &= -\D{\mathcal{D}(\omega,\mvec{k})}{t} = 0,\qquad
\Rightarrow\qquad \omega = \mathrm{const}\\ 
\dd{r_l}{\tau} &= \D{\mathcal{D}(\omega,\mvec{k})}{k_l} = 2\omega k_l + \partial_i \Sigma_{il},\\
\dd{k_l}{\tau} &= -\D{\mathcal{D}(\omega,\mvec{k})}{r_l} =  -\partial_l\partial_i \Sigma_{ij} k_j.
\end{align}
We integrate these equations
starting at $\tau$=0, $t=0$, $\mvec{r}=\mvec{r}_0$ and $\mvec{k}=\mvec{k}_0$ to 
trace the wave trajectory $t(\tau)$, $\mvec{r}(\tau)$ and $\mvec{k}(\tau)$.

For numerical integration it is beneficial to split $\mvec{k}$ into
magnitude $k$ and direction $\tilde{\mvec{k}}$ parts ($\mvec{k} = k
\mvec{\tilde{k}}$, and $|\tilde{\mvec{k}}|^2= 1$), and rewrite the
last two equations as  
\begin{align}
\label{eq::ray_t}
\dd{r_l}{\tau} &= -2\partial_i\Sigma_{ij} \tilde{k}_j\tilde{k}_l + \partial_i \Sigma_{il},\\
\dd{\tilde{k}_l}{\tau} &= -\partial_l\partial_i \Sigma_{ij} \tilde{k}_j +
\partial_m\partial_i \Sigma_{ij} \tilde{k}_j\tilde{k}_m\tilde{k}_l,
\end{align}
where in the last equation the right hand side is orthogonal to
$\tilde{\mvec{k}}$ to guarantee that $|\tilde{\mvec{k}}|^2= 1$, and
the algebraic expression \cref{eq::dispI} was substituted for the
differential equation for wave vector magnitude $k$. 

\section{Wave dissipation and excitation}
\label{sec:spiking}
An integral along the wave trajectory $t(\tau)$, $\mvec{r}(\tau)$ and $\mvec{k}(\tau)$
\begin{align}
\label{eq::energy}
W &= W_0\exp\left(-\int\limits_0
\left[\gamma_{dis}\left(\mvec{r}(\tau)\right) -
  \gamma_{exc}\left(\mvec{r}(\tau)\right)\right]dt\right) \n 
 &= W_0\exp\left(-\int\limits_0
\left[\Sigma_{ij}\left(\mvec{r}(\tau)\right)\tilde{k}_i(\tau)\tilde{k}_j(\tau)
  - \gamma_{exc}\left(\mvec{r}(\tau)\right)\right]k^2(\tau)
d\tau\right)  
\end{align}
with any appropriate model form of wave excitation $\gamma_{exc}$
describes a change of wave energy $W$ along its path. This may allow
the study of many interesting questions of brain wave
dynamics, i.e.~to identify potential active area where wave intensity
may grow as a result of certain frequency and/or spatial distributions
of neuronal spiky activation, or to estimate the spatial
extent of coherent activation area as a result of some particular
point sources, or to find when and/or how these waves may potentially
become important in triggering neuronal firing, that is to study
possible mechanisms of synchronization and feedback, etc.

\section{Persistent wave loop patterns}

One of the interesting questions is 
the possibility of persistent pattern
formation/self\-organization from a randomly emitted distribution of
brain waves influenced by inhomogeous anisotropic structure of their
dispersion. Considering a simplest case of fixed difference between
wave excitation and dissipation, i.e.~assuming propagation of wave
packets of fixed width (exponentially decaying) and searching for any
closed parts of wave trajectories (loops) that are shorter than the
packet width can provide a clue about possible answer. \Cref{fig:rays}
as well as \crefrange{S1}{S16} (and hyperlinked movies) show
formation of persistent loops for a variety of wave packet initial
conditions as well as for different models of conductivity tensor
inhomogeneity and anisotropy constructed from dMRI or whole brain
tractography.

Examples of wave trajectories
and emergent persistent loop patterns for the spherical shell cortex
model with varying amounts of tensor anisotropy and inhomogeneous
shell layer thickness (\crefrange{S1}{S4}).  

Examples of wave trajectories
and emergent persistent loop patterns for cortical fold geometry with
different approaches used for estimation of inhomogeneity and
anisotropy are shown in \crefrange{S5}{S16}. Among those examples are
several simple cases with variable inhomogeneity and fixed anisotropy
(similar to the above spherical shell cortex model) as well as with
more complex estimates of anisotropy based on multiple diffusion
gradients MRI (dMRI) acquisitions. Assuming linear dependence between
diffusion and conductivity tensors \cite{pmid11573005} several
combination of the diffusion tensor eigenvectors were used to describe
the conductivity tensor anisotropy. In even more complex approach, the
full brain tractography \cite{Galinsky:2015} generated fiber distributions
\cite{brain-grid} were used to infer anisotropy through direct
integration of single fiber anisotropy parameters in each voxel.

All wave trajectory figures are hyperlinked and include references
to location of movies showing dynamical development of wave trajectories.

\clearpage
\begin{figure}
\centering
\begin{tikzpicture}[scale=\fsc]%
T\pgfoonew \mt=new tmplt()
\mt.apply(\Ref{S01}-2-s.mp4,\Ref{S01}-b.png,\Ref{S01c0},\Ref{S01}-1-s.mp4,\Ref{S01}-e.png,\Ref{S01c1}) 
\end{tikzpicture}
\caption{An example of complete wave trajectory (top) and emergent
  loop pattern (bottom) for the spherical shell cortex model with
  crossing fibers anisotropy tensor $\sigma^{(3)}$ and narrow
  inhomogeneity layer ($n$=0, $r_0$=0.5 and $r_1$=0.9).  The
  trajectory was initialized with wave vector
  $\mvec{k}=(-0.1/\sqrt{2}, 0.1/\sqrt{2}, -0.1)$ inside the
  inhomogeneous layer with voxel coordinates $\mvec{r}$=(142,142,142).
  High resolution movie links:
  \href{\Ref{S01c0}}{S1-H1}/\href{\Ref{SA01c0}}{S1-H1},
  \href{\Ref{S01c1}}{S1-H2}/\href{\Ref{SA01c1}}{S1-H2}.  Low
  resolution movie links:
  \href{\Ref{SS01c0}}{S1-L1}/\href{\Ref{SSA01c0}}{S1-L1},
  \href{\Ref{SS01c1}}{S1-L2}/\href{\Ref{SSA01c1}}{S1-L2}.
\label{S1}
}
\end{figure}

\begin{figure}
\centering
\begin{tikzpicture}[scale=\fsc]%
\pgfoonew \mt=new tmplt()
\mt.apply(\Ref{S02}-1-s.mp4,\Ref{S02}-b.png,\Ref{S02c0},\Ref{S02}-2-s.mp4,\Ref{S02}-e.png,\Ref{S02c1})
\end{tikzpicture}
\caption{ An example of complete wave trajectory (top) and emergent
  loop pattern (bottom) for the spherical shell cortex model with
  crossing fibers anisotropy tensor $\sigma^{(3)}$ and slightly wider
  inhomogeneity layer ($n$=2, $r_0$=0.5 and $r_1$=0.9).  The
  trajectory was initialized with wave vector
  $\mvec{k}=(-0.1/\sqrt{2}, 0.1/\sqrt{2}, -0.1)$ inside the
  inhomogeneous layer with voxel coordinates $\mvec{r}$=(80,33,100).
  High resolution movie links:
  \href{\Ref{S02c0}}{S2-H1}/\href{\Ref{SA02c0}}{S2-H1},
  \href{\Ref{S02c1}}{S2-H2}/\href{\Ref{SA02c1}}{S2-H2}.  Low
  resolution movie links:
  \href{\Ref{SS02c0}}{S2-L1}/\href{\Ref{SSA02c0}}{S2-L1},
  \href{\Ref{SS02c1}}{S2-L2}/\href{\Ref{SSA02c1}}{S2-L2}.  
\label{S2}
}
\end{figure}

\begin{figure}
\centering
\begin{tikzpicture}[scale=\fsc]%
\pgfoonew \mt=new tmplt()
\mt.apply(\Ref{S03}-1-s.mp4,\Ref{S03}-b.png,\Ref{S03c0},\Ref{S03}-2-s.mp4,\Ref{S03}-e.png,\Ref{S03c1})
\end{tikzpicture}
\caption{ An example of complete wave trajectory (top) and emergent
  loop pattern (bottom) for the spherical shell cortex model with
  45$^\circ$ orientation single fiber anisotropy tensor $\sigma^{(2)}$
  and wide inhomogeneity layer ($n$=25, $r_0$=0.5 and $r_1$=0.9).  The
  trajectory was initialized with wave vector
  $\mvec{k}=(-0.1/\sqrt{2}, 0.1/\sqrt{2}, -0.1)$ inside the
  inhomogeneous layer with voxel coordinates $\mvec{r}$=(145,145,145).
  High resolution movie links:
  \href{\Ref{S03c0}}{S3-H1}/\href{\Ref{SA03c0}}{S3-H1},
  \href{\Ref{S03c1}}{S3-H2}/\href{\Ref{SA03c1}}{S3-H2}.  Low
  resolution movie links:
  \href{\Ref{SS03c0}}{S3-L1}/\href{\Ref{SSA03c0}}{S3-L1},
  \href{\Ref{SS03c1}}{S3-L2}/\href{\Ref{SSA03c1}}{S3-L2}.  
\label{S3}
}
\end{figure}

\begin{figure}
\centering
\begin{tikzpicture}[scale=\fsc]%
\pgfoonew \mt=new tmplt()
\mt.apply(\Ref{S04}-1-s.mp4,\Ref{S04}-b.png,\Ref{S04c0},\Ref{S04}-2-s.mp4,\Ref{S04}-e.png,\Ref{S04c1})
\end{tikzpicture}
\caption{ An example of complete wave trajectory (top) and emergent
  loop pattern (bottom) for the spherical shell cortex model with
  45$^\circ$ orientation single fiber anisotropy tensor $\sigma^{(2)}$
  and intermediately wide inhomogeneity layer ($n$=14, $r_0$=0.5 and
  $r_1$=0.9).  The trajectory was initialized with wave vector
  $\mvec{k}=(-0.1/\sqrt{2}, 0.1/\sqrt{2}, -0.1)$ inside the
  inhomogeneous layer with voxel coordinates $\mvec{r}$=(141,141,141).
  High resolution movie links:
  \href{\Ref{S04c0}}{S4-H1}/\href{\Ref{SA04c0}}{S4-H1},
  \href{\Ref{S04c1}}{S4-H2}/\href{\Ref{SA04c1}}{S4-H2}.  Low
  resolution movie links:
  \href{\Ref{SS04c0}}{S4-L1}/\href{\Ref{SSA04c0}}{S4-L1},
  \href{\Ref{SS04c1}}{S4-L2}/\href{\Ref{SSA04c1}}{S4-L2}.  
\label{S4}
}
\end{figure}

\begin{figure}
\centering
\begin{tikzpicture}[scale=\fsc]%
\pgfoonew \mt=new tmplt()
\mt.apply(\Ref{S05}-3-s.mp4,\Ref{S05}-b.png,\Ref{S05c0},\Ref{S05}-1-s.mp4,\Ref{S05}-e.png,\Ref{S05c1})
\end{tikzpicture}
\caption{ An example of randomly initialized complete wave trajectory
  (top) and emergent loop pattern (bottom) for the cortical fold model
  with inhomogeneity extracted from HRA volume registered to MNI152
  space and with diffusion MRI derived position dependent anisotropy
  tensor $\mvec{R}^T \mvec{\sigma}^{(1)}\mvec{R}$
  ($\varepsilon=(\lambda_d^{(2)}+\lambda_d^{(3)})/2/\lambda_d^{(1)}$).
  The trajectory was initialized with wave vector $\mvec{k}$=(-0.75,
  0.23, -0.62) inside the inhomogeneous layer with voxel coordinates
  $\mvec{r}$=(55,130,130).  High resolution movie links:
  \href{\Ref{S05c0}}{S5-H1}/\href{\Ref{SA05c0}}{S5-H1},
  \href{\Ref{S05c1}}{S5-H2}/\href{\Ref{SA05c1}}{S5-H2},
  \href{\Ref{S05c2}}{S5-H3}/\href{\Ref{SA05c2}}{S5-H3}.  Low
  resolution movie links:
  \href{\Ref{SS05c0}}{S5-L1}/\href{\Ref{SSA05c0}}{S5-L1},
  \href{\Ref{SS05c1}}{S5-L2}/\href{\Ref{SSA05c1}}{S5-L2},
  \href{\Ref{SS05c2}}{S5-L3}/\href{\Ref{SSA05c2}}{S5-L3}.  
\label{S5}
}
\end{figure}

\begin{figure}
\centering
\begin{tikzpicture}[scale=\fsc]%
\pgfoonew \mt=new tmplt()
\mt.apply(\Ref{S06}-3-s.mp4,\Ref{S06}-b.png,\Ref{S06c0},\Ref{S06}-1-s.mp4,\Ref{S06}-e.png,\Ref{S06c1})
\end{tikzpicture}
\caption{ An example of randomly initialized complete wave trajectory
  (top) and emergent loop pattern (bottom) for the cortical fold model
  with inhomogeneity extracted from HRA volume registered to MNI152
  space and with diffusion MRI derived position dependent anisotropy
  tensor $\mvec{R}^T \mvec{\sigma}^{(1)}\mvec{R}$
  ($\varepsilon=\lambda_d^{(2)}/\lambda_d^{(1)}$).  The trajectory was
  initialized with wave vector $\mvec{k}$=(-0.9,0.24,0.36) inside the
  inhomogeneous layer with voxel coordinates $\mvec{r}$=(66,94,126).
  High resolution movie links:
  \href{\Ref{S06c0}}{S6-H1}/\href{\Ref{SA06c0}}{S6-H1},
  \href{\Ref{S06c1}}{S6-H2}/\href{\Ref{SA06c1}}{S6-H2},
  \href{\Ref{S06c2}}{S6-H3}/\href{\Ref{SA06c2}}{S6-H3}.  Low
  resolution movie links:
  \href{\Ref{SS06c0}}{S6-L1}/\href{\Ref{SSA06c0}}{S6-L1},
  \href{\Ref{SS06c1}}{S6-L2}/\href{\Ref{SSA06c1}}{S6-L2},
  \href{\Ref{SS06c2}}{S6-L3}/\href{\Ref{SSA06c2}}{S6-L3}.  
\label{S6}
}
\end{figure}

\begin{figure}
\centering
\begin{tikzpicture}[scale=\fsc]%
\pgfoonew \mt=new tmplt()
\mt.apply(\Ref{S07}-3-s.mp4,\Ref{S07}-b.png,\Ref{S07c0},\Ref{S07}-1-s.mp4,\Ref{S07}-e.png,\Ref{S07c1})
\end{tikzpicture}
\caption{ An example of randomly initialized complete wave trajectory
  (top) and emergent loop pattern (bottom) for the cortical fold model
  with inhomogeneity extracted from HRA volume registered to MNI152
  space and with diffusion MRI derived position dependent anisotropy
  tensor $\mvec{R}^T \mvec{\sigma}^{(1)}\mvec{R}$
  ($\sigma^{(1)}_{22}=\lambda_d^{(2)}/\lambda_d^{(1)}$ and
  $\sigma^{(1)}_{11}=\lambda_d^{(3)}/\lambda_d^{(1)})$.  The
  trajectory was initialized with wave vector
  $\mvec{k}$=(-0.77,-0.63,0.11) inside the inhomogeneous layer with
  voxel coordinates $\mvec{r}$=(49,153,97).  High resolution movie
  links: \href{\Ref{S07c0}}{S7-H1}/\href{\Ref{SA07c0}}{S7-H1},
  \href{\Ref{S07c1}}{S7-H2}/\href{\Ref{SA07c1}}{S7-H2},
  \href{\Ref{S07c2}}{S7-H3}/\href{\Ref{SA07c2}}{S7-H3}.  Low
  resolution movie links:
  \href{\Ref{SS07c0}}{S7-L1}/\href{\Ref{SSA07c0}}{S7-L1},
  \href{\Ref{SS07c1}}{S7-L2}/\href{\Ref{SSA07c1}}{S7-L2},
  \href{\Ref{SS07c2}}{S7-L3}/\href{\Ref{SSA07c2}}{S7-L3}.  
\label{S7}
}
\end{figure}

\begin{figure}
\centering
\begin{tikzpicture}[scale=\fsc]%
\pgfoonew \mt=new tmplt()
\mt.apply(\Ref{S08}-3-s.mp4,\Ref{S08}-b.png,\Ref{S08c0},\Ref{S08}-1-s.mp4,\Ref{S08}-e.png,\Ref{S08c1})
\end{tikzpicture}
\caption{ An example of randomly initialized complete wave trajectory
  (top) and emergent loop pattern (bottom) for the cortical fold model
  with inhomogeneity extracted from HRA volume registered to MNI152
  space and with fixed anisotropy tensor $\sigma^{(1)}$
  ($\varepsilon=0.1$).  The trajectory was initialized with wave
  vector $\mvec{k}$=(0.56,0.58,-0.59) inside the inhomogeneous layer
  with voxel coordinates $\mvec{r}$=(72,156,120).  High resolution
  movie links: \href{\Ref{S08c0}}{S8-H1}/\href{\Ref{SA08c0}}{S8-H1},
  \href{\Ref{S08c1}}{S8-H2}/\href{\Ref{SA08c1}}{S8-H2},
  \href{\Ref{S08c2}}{S8-H3}/\href{\Ref{SA08c2}}{S8-H3}.  Low
  resolution movie links:
  \href{\Ref{SS08c0}}{S8-L1}/\href{\Ref{SSA08c0}}{S8-L1},
  \href{\Ref{SS08c1}}{S8-L2}/\href{\Ref{SSA08c1}}{S8-L2},
  \href{\Ref{SS08c2}}{S8-L3}/\href{\Ref{SSA08c2}}{S8-L3}.  
\label{S8}
}
\end{figure}

\begin{figure}
\centering
\begin{tikzpicture}[scale=\fsc]%
\pgfoonew \mt=new tmplt()
\mt.apply(\Ref{S09}-3-s.mp4,\Ref{S09}-b.png,\Ref{S09c0},\Ref{S09}-1-s.mp4,\Ref{S09}-e.png,\Ref{S09c1})
\end{tikzpicture}
\caption{ An example of randomly initialized complete wave trajectory
  (top) and emergent loop pattern (bottom) for the cortical fold model
  with inhomogeneity extracted from HRA volume registered to MNI152
  space and with diffusion MRI derived position independent anisotropy
  tensor $\mvec{R}^T \mvec{\sigma}^{(1)}\mvec{R}$ ($\varepsilon=0.1$).
  The trajectory was initialized with wave vector
  $\mvec{k}$=(-0.13,-0.97,-0.21) inside the inhomogeneous layer with
  voxel coordinates $\mvec{r}$=(111,103,110).  High resolution movie
  links: \href{\Ref{S09c0}}{S9-H1}/\href{\Ref{SA09c0}}{S9-H1},
  \href{\Ref{S09c1}}{S9-H2}/\href{\Ref{SA09c1}}{S9-H2},
  \href{\Ref{S09c2}}{S9-H3}/\href{\Ref{SA09c2}}{S9-H3}.  Low
  resolution movie links:
  \href{\Ref{SS09c0}}{S9-L1}/\href{\Ref{SSA09c0}}{S9-L1},
  \href{\Ref{SS09c1}}{S9-L2}/\href{\Ref{SSA09c1}}{S9-L2},
  \href{\Ref{SS09c2}}{S9-L3}/\href{\Ref{SSA09c2}}{S9-L3}.  
\label{S9}
}
\end{figure}

\begin{figure}
\centering
\begin{tikzpicture}[scale=\fsc]%
\pgfoonew \mt=new tmplt()
\mt.apply(\Ref{S10}-3-s.mp4,\Ref{S10}-b.png,\Ref{S10c0},\Ref{S10}-1-s.mp4,\Ref{S10}-e.png,\Ref{S10c1})
\end{tikzpicture}
\caption{ An example of randomly initialized complete wave trajectory
  (top) and emergent loop pattern (bottom) for the cortical fold model
  with inhomogeneity extracted from HRA volume registered to MNI152
  space and with fixed anisotropy tensor $\sigma^{(1)}$
  ($\varepsilon=0.01$).  The trajectory was initialized with wave
  vector $\mvec{k}$=(0.42,-0.41,0.81) inside the inhomogeneous layer
  with voxel coordinates $\mvec{r}$=(114,85,22).  High resolution
  movie links: \href{\Ref{S10c0}}{S10-H1}/\href{\Ref{SA10c0}}{S10-H1},
  \href{\Ref{S10c1}}{S10-H2}/\href{\Ref{SA10c1}}{S10-H2},
  \href{\Ref{S10c2}}{S10-H3}/\href{\Ref{SA10c2}}{S10-H3}.  Low
  resolution movie links:
  \href{\Ref{SS10c0}}{S10-L1}/\href{\Ref{SSA10c0}}{S10-L1},
  \href{\Ref{SS10c1}}{S10-L2}/\href{\Ref{SSA10c1}}{S10-L2},
  \href{\Ref{SS10c2}}{S10-L3}/\href{\Ref{SSA10c2}}{S10-L3}.  
\label{S10}
}
\end{figure}

\begin{figure}
\centering
\begin{tikzpicture}[scale=\fsc]%
\pgfoonew \mt=new tmplt()
\mt.apply(\Ref{S11}-3-s.mp4,\Ref{S11}-b.png,\Ref{S11c0},\Ref{S11}-1-s.mp4,\Ref{S11}-e.png,\Ref{S11c1})
\end{tikzpicture}
\caption{ An example of randomly initialized complete wave trajectory
  (top) and emergent loop pattern (bottom) for the cortical fold model
  with inhomogeneity extracted from HRA volume registered to MNI152
  space and with fixed anisotropy tensor $\sigma^{(2)}$
  ($\varepsilon=0.1$).  The trajectory was initialized with wave
  vector $\mvec{k}$=(-0.90,0.24,0.36) inside the inhomogeneous layer
  with voxel coordinates $\mvec{r}$=(66,86,126).  High resolution
  movie links: \href{\Ref{S11c0}}{S11-H1}/\href{\Ref{SA11c0}}{S11-H1},
  \href{\Ref{S11c1}}{S11-H2}/\href{\Ref{SA11c1}}{S11-H2},
  \href{\Ref{S11c2}}{S11-H3}/\href{\Ref{SA11c2}}{S11-H3}.  Low
  resolution movie links:
  \href{\Ref{SS11c0}}{S11-L1}/\href{\Ref{SSA11c0}}{S11-L1},
  \href{\Ref{SS11c1}}{S11-L2}/\href{\Ref{SSA11c1}}{S11-L2},
  \href{\Ref{SS11c2}}{S11-L3}/\href{\Ref{SSA11c2}}{S11-L3}.  
\label{S11}
}
\end{figure}

\begin{figure}
\centering
\begin{tikzpicture}[scale=\fsc]%
\pgfoonew \mt=new tmplt()
\mt.apply(\Ref{S12}-3-s.mp4,\Ref{S12}-b.png,\Ref{S12c0},\Ref{S12}-1-s.mp4,\Ref{S12}-e.png,\Ref{S12c1})
\end{tikzpicture}
\caption{ An example of randomly initialized complete wave trajectory
  (top) and emergent loop pattern (bottom) for the cortical fold model
  with inhomogeneity extracted from HRA volume registered to MNI152
  space and with diffusion MRI derived position independent anisotropy
  tensor $\mvec{R}^T \mvec{\sigma}^{(1)}\mvec{R}$
  ($\varepsilon=0.01$).  The trajectory was initialized with wave
  vector $\mvec{k}$=(-0.90,0.24,0.36) inside the inhomogeneous layer
  with voxel coordinates $\mvec{r}$=(66,87,126).  High resolution
  movie links: \href{\Ref{S12c0}}{S12-H1}/\href{\Ref{SA12c0}}{S12-H1},
  \href{\Ref{S12c1}}{S12-H2}/\href{\Ref{SA12c1}}{S12-H2},
  \href{\Ref{S12c2}}{S12-H3}/\href{\Ref{SA12c2}}{S12-H3}.  Low
  resolution movie links:
  \href{\Ref{SS12c0}}{S12-L1}/\href{\Ref{SSA12c0}}{S12-L1},
  \href{\Ref{SS12c1}}{S12-L2}/\href{\Ref{SSA12c1}}{S12-L2},
  \href{\Ref{SS12c2}}{S12-L3}/\href{\Ref{SSA12c2}}{S12-L3}.  
\label{S12}
}
\end{figure}

\begin{figure}
\centering
\begin{tikzpicture}[scale=\fsc]%
\pgfoonew \mt=new tmplt()
\mt.apply(\Ref{S13}-3-s.mp4,\Ref{S13}-b.png,\Ref{S13c0},\Ref{S13}-1-s.mp4,\Ref{S13}-e.png,\Ref{S13c1})
\end{tikzpicture}
\caption{ An example of randomly initialized complete wave trajectory
  (top) and emergent loop pattern (bottom) for the cortical fold model
  with inhomogeneity extracted from HRA volume registered to MNI152
  space and with tractography derived anisotropy tensor
  $1/N\sum_k\mvec{R}_k^T \mvec{\sigma}^{(1)}\mvec{R}_k$
  ($\varepsilon=0.001$).  The trajectory was initialized with wave
  vector $\mvec{k}$=(0.4,-0.83,0.4) inside the inhomogeneous layer
  with voxel coordinates $\mvec{r}$=(75,61,90).  High resolution movie
  links: \href{\Ref{S13c0}}{S13-H1}/\href{\Ref{SA13c0}}{S13-H1},
  \href{\Ref{S13c1}}{S13-H2}/\href{\Ref{SA13c1}}{S13-H2},
  \href{\Ref{S13c2}}{S13-H3}/\href{\Ref{SA13c2}}{S13-H3}.  Low
  resolution movie links:
  \href{\Ref{SS13c0}}{S13-L1}/\href{\Ref{SSA13c0}}{S13-L1},
  \href{\Ref{SS13c1}}{S13-L2}/\href{\Ref{SSA13c1}}{S13-L2},
  \href{\Ref{SS13c2}}{S13-L3}/\href{\Ref{SSA13c2}}{S13-L3}.  
\label{S13}
}
\end{figure}

\begin{figure}
\centering
\begin{tikzpicture}[scale=\fsc]%
\pgfoonew \mt=new tmplt()
\mt.apply(\Ref{S14}-3-s.mp4,\Ref{S14}-b.png,\Ref{S14c0},\Ref{S14}-1-s.mp4,\Ref{S14}-e.png,\Ref{S14c1})
\end{tikzpicture}
\caption{ An example of randomly initialized complete wave trajectory
  (top) and emergent loop pattern (bottom) for the cortical fold model
  with inhomogeneity extracted from HRA volume registered to MNI152
  space and with tractography derived anisotropy tensor
  $1/N\sum_k\mvec{R}_k^T \mvec{\sigma}^{(1)}\mvec{R}_k$
  ($\varepsilon=0.01$).  The trajectory was initialized with wave
  vector $\mvec{k}$=(0.66,0.5,0.56) inside the inhomogeneous layer
  with voxel coordinates $\mvec{r}$=(33,77,56).  High resolution movie
  links: \href{\Ref{S14c0}}{S14-H1}/\href{\Ref{SA14c0}}{S14-H1},
  \href{\Ref{S14c1}}{S14-H2}/\href{\Ref{SA14c1}}{S14-H2},
  \href{\Ref{S14c2}}{S14-H3}/\href{\Ref{SA14c2}}{S14-H3}.  Low
  resolution movie links:
  \href{\Ref{SS14c0}}{S14-L1}/\href{\Ref{SSA14c0}}{S14-L1},
  \href{\Ref{SS14c1}}{S14-L2}/\href{\Ref{SSA14c1}}{S14-L2},
  \href{\Ref{SS14c2}}{S14-L3}/\href{\Ref{SSA14c2}}{S14-L3}.  
\label{S14}
}
\end{figure}

\begin{figure}
\centering
\begin{tikzpicture}[scale=\fsc]%
\pgfoonew \mt=new tmplt()
\mt.apply(\Ref{S15}-3-s.mp4,\Ref{S15}-b.png,\Ref{S15c0},\Ref{S15}-1-s.mp4,\Ref{S15}-e.png,\Ref{S15c1})
\end{tikzpicture}
\caption{ An example of randomly initialized complete wave trajectory
  (top) and emergent loop pattern (bottom) for the cortical fold model
  with inhomogeneity extracted from HRA volume registered to MNI152
  space and with tractography derived anisotropy tensor
  $1/N\sum_k\mvec{R}_k^T \mvec{\sigma}^{(1)}\mvec{R}_k$
  ($\varepsilon=0.1$).  The trajectory was initialized with wave
  vector $\mvec{k}$=(0.36,0.67,-0.65) inside the inhomogeneous layer
  with voxel coordinates $\mvec{r}$=(77,123,64).  High resolution
  movie links: \href{\Ref{S15c0}}{S15-H1}/\href{\Ref{SA15c0}}{S15-H1},
  \href{\Ref{S15c1}}{S15-H2}/\href{\Ref{SA15c1}}{S15-H2},
  \href{\Ref{S15c2}}{S15-H3}/\href{\Ref{SA15c2}}{S15-H3}.  Low
  resolution movie links:
  \href{\Ref{SS15c0}}{S15-L1}/\href{\Ref{SSA15c0}}{S15-L1},
  \href{\Ref{SS15c1}}{S15-L2}/\href{\Ref{SSA15c1}}{S15-L2},
  \href{\Ref{SS15c2}}{S15-L3}/\href{\Ref{SSA15c2}}{S15-L3}.  
\label{S15}
}
\end{figure}

\begin{figure}
\centering
\begin{tikzpicture}[scale=\fsc]%
\pgfoonew \mt=new tmplt()
\mt.apply(\Ref{S16}-3-s.mp4,\Ref{S16}-b.png,\Ref{S16c0},\Ref{S16}-1-s.mp4,\Ref{S16}-e.png,\Ref{S16c1})
\end{tikzpicture}
\caption{ An example of randomly initialized complete wave trajectory
  (top) and emergent loop pattern (bottom) for the cortical fold model
  with inhomogeneity extracted from HRA volume registered to MNI152
  space and with tractography derived anisotropy tensor
  $1/N\sum_k\mvec{R}_k^T \mvec{\sigma}^{(1)}\mvec{R}_k$
  ($\varepsilon=0$).  The trajectory was initialized with wave vector
  $\mvec{k}$=(-0.06,0.38,0.92) inside the inhomogeneous layer with
  voxel coordinates $\mvec{r}$=(132,113,96).  High resolution movie
  links: \href{\Ref{S16c0}}{S16-H1}/\href{\Ref{SA16c0}}{S16-H1},
  \href{\Ref{S16c1}}{S16-H2}/\href{\Ref{SA16c1}}{S16-H2},
  \href{\Ref{S16c2}}{S16-H3}/\href{\Ref{SA16c2}}{S16-H3}.  Low
  resolution movie links:
  \href{\Ref{SS16c0}}{S16-L1}/\href{\Ref{SSA16c0}}{S16-L1},
  \href{\Ref{SS16c1}}{S16-L2}/\href{\Ref{SSA16c1}}{S16-L2},
  \href{\Ref{SS16c2}}{S16-L3}/\href{\Ref{SSA16c2}}{S16-L3}.  
\label{S16}
}
\end{figure}

\vfill

\clearpage
\twocolumngrid


\begin{thebibliography}{39}%
\makeatletter
\providecommand \@ifxundefined [1]{%
 \@ifx{#1\undefined}
}%
\providecommand \@ifnum [1]{%
 \ifnum #1\expandafter \@firstoftwo
 \else \expandafter \@secondoftwo
 \fi
}%
\providecommand \@ifx [1]{%
 \ifx #1\expandafter \@firstoftwo
 \else \expandafter \@secondoftwo
 \fi
}%
\providecommand \natexlab [1]{#1}%
\providecommand \enquote  [1]{``#1''}%
\providecommand \bibnamefont  [1]{#1}%
\providecommand \bibfnamefont [1]{#1}%
\providecommand \citenamefont [1]{#1}%
\providecommand \href@noop [0]{\@secondoftwo}%
\providecommand \href [0]{\begingroup \@sanitize@url \@href}%
\providecommand \@href[1]{\@@startlink{#1}\@@href}%
\providecommand \@@href[1]{\endgroup#1\@@endlink}%
\providecommand \@sanitize@url [0]{\catcode `\\12\catcode `\$12\catcode
  `\&12\catcode `\#12\catcode `\^12\catcode `\_12\catcode `\%12\relax}%
\providecommand \@@startlink[1]{}%
\providecommand \@@endlink[0]{}%
\providecommand \url  [0]{\begingroup\@sanitize@url \@url }%
\providecommand \@url [1]{\endgroup\@href {#1}{\urlprefix }}%
\providecommand \urlprefix  [0]{URL }%
\providecommand \Eprint [0]{\href }%
\providecommand \doibase [0]{https://doi.org/}%
\providecommand \selectlanguage [0]{\@gobble}%
\providecommand \bibinfo  [0]{\@secondoftwo}%
\providecommand \bibfield  [0]{\@secondoftwo}%
\providecommand \translation [1]{[#1]}%
\providecommand \BibitemOpen [0]{}%
\providecommand \bibitemStop [0]{}%
\providecommand \bibitemNoStop [0]{.\EOS\space}%
\providecommand \EOS [0]{\spacefactor3000\relax}%
\providecommand \BibitemShut  [1]{\csname bibitem#1\endcsname}%
\let\auto@bib@innerbib\@empty
\bibitem [{\citenamefont {Monai}\ \emph {et~al.}(2012)\citenamefont {Monai},
  \citenamefont {Inoue}, \citenamefont {Miyakawa},\ and\ \citenamefont
  {Aonishi}}]{pmid23367980}%
  \BibitemOpen
  \bibfield  {author} {\bibinfo {author} {\bibfnamefont {H.}~\bibnamefont
  {Monai}}, \bibinfo {author} {\bibfnamefont {M.}~\bibnamefont {Inoue}},
  \bibinfo {author} {\bibfnamefont {H.}~\bibnamefont {Miyakawa}},\ and\
  \bibinfo {author} {\bibfnamefont {T.}~\bibnamefont {Aonishi}},\ }\bibfield
  {title} {\bibinfo {title} {{{L}ow-frequency dielectric dispersion of brain
  tissue due to electrically long neurites}},\ }\href@noop {} {\bibfield
  {journal} {\bibinfo  {journal} {Phys Rev E Stat Nonlin Soft Matter Phys}\
  }\textbf {\bibinfo {volume} {86}},\ \bibinfo {pages} {061911} (\bibinfo
  {year} {2012})}\BibitemShut {NoStop}%
\bibitem [{\citenamefont {Ingber}\ and\ \citenamefont
  {Nunez}(2011)}]{pmid21167841}%
  \BibitemOpen
  \bibfield  {author} {\bibinfo {author} {\bibfnamefont {L.}~\bibnamefont
  {Ingber}}\ and\ \bibinfo {author} {\bibfnamefont {P.~L.}\ \bibnamefont
  {Nunez}},\ }\bibfield  {title} {\bibinfo {title} {{{N}eocortical dynamics at
  multiple scales: {E}{E}{G} standing waves, statistical mechanics, and
  physical analogs}},\ }\href@noop {} {\bibfield  {journal} {\bibinfo
  {journal} {Math Biosci}\ }\textbf {\bibinfo {volume} {229}},\ \bibinfo
  {pages} {160} (\bibinfo {year} {2011})}\BibitemShut {NoStop}%
\bibitem [{\citenamefont {{Li}}\ \emph {et~al.}(2014)\citenamefont {{Li}},
  \citenamefont {{Xia}}, \citenamefont {{Qu}}, \citenamefont {{Wu}},
  \citenamefont {{Yang}}, \citenamefont {{Hao}}, \citenamefont {{Jiang}},\ and\
  \citenamefont {{Li}}}]{2014NatSR...4E6893L}%
  \BibitemOpen
  \bibfield  {author} {\bibinfo {author} {\bibfnamefont {X.~P.}\ \bibnamefont
  {{Li}}}, \bibinfo {author} {\bibfnamefont {Q.}~\bibnamefont {{Xia}}},
  \bibinfo {author} {\bibfnamefont {D.}~\bibnamefont {{Qu}}}, \bibinfo {author}
  {\bibfnamefont {T.~C.}\ \bibnamefont {{Wu}}}, \bibinfo {author}
  {\bibfnamefont {D.~G.}\ \bibnamefont {{Yang}}}, \bibinfo {author}
  {\bibfnamefont {W.~D.}\ \bibnamefont {{Hao}}}, \bibinfo {author}
  {\bibfnamefont {X.}~\bibnamefont {{Jiang}}},\ and\ \bibinfo {author}
  {\bibfnamefont {X.~M.}\ \bibnamefont {{Li}}},\ }\bibfield  {title} {\bibinfo
  {title} {{The Dynamic Dielectric at a Brain Functional Site and an EM Wave
  Approach to Functional Brain Imaging}},\ }\href
  {https://doi.org/10.1038/srep06893} {\bibfield  {journal} {\bibinfo
  {journal} {Scientific Reports}\ }\textbf {\bibinfo {volume} {4}},\ \bibinfo
  {eid} {6893} (\bibinfo {year} {2014})}\BibitemShut {NoStop}%
\bibitem [{\citenamefont {Zalesky}\ \emph {et~al.}(2014)\citenamefont
  {Zalesky}, \citenamefont {Fornito}, \citenamefont {Cocchi}, \citenamefont
  {Gollo},\ and\ \citenamefont {Breakspear}}]{pmid24982140}%
  \BibitemOpen
  \bibfield  {author} {\bibinfo {author} {\bibfnamefont {A.}~\bibnamefont
  {Zalesky}}, \bibinfo {author} {\bibfnamefont {A.}~\bibnamefont {Fornito}},
  \bibinfo {author} {\bibfnamefont {L.}~\bibnamefont {Cocchi}}, \bibinfo
  {author} {\bibfnamefont {L.~L.}\ \bibnamefont {Gollo}},\ and\ \bibinfo
  {author} {\bibfnamefont {M.}~\bibnamefont {Breakspear}},\ }\bibfield  {title}
  {\bibinfo {title} {{{T}ime-resolved resting-state brain networks}},\
  }\href@noop {} {\bibfield  {journal} {\bibinfo  {journal} {Proc. Natl. Acad.
  Sci. U.S.A.}\ }\textbf {\bibinfo {volume} {111}},\ \bibinfo {pages} {10341}
  (\bibinfo {year} {2014})}\BibitemShut {NoStop}%
\bibitem [{\citenamefont {Van~Essen}\ \emph {et~al.}(2013)\citenamefont
  {Van~Essen}, \citenamefont {Smith}, \citenamefont {Barch}, \citenamefont
  {Behrens}, \citenamefont {Yacoub}, \citenamefont {Ugurbil},\ and\
  \citenamefont {for~the WU-Minn HCP~Consortium}}]{VanEssen:2013}%
  \BibitemOpen
  \bibfield  {author} {\bibinfo {author} {\bibfnamefont {D.~C.}\ \bibnamefont
  {Van~Essen}}, \bibinfo {author} {\bibfnamefont {S.~M.}\ \bibnamefont
  {Smith}}, \bibinfo {author} {\bibfnamefont {D.~M.}\ \bibnamefont {Barch}},
  \bibinfo {author} {\bibfnamefont {T.~E.~J.}\ \bibnamefont {Behrens}},
  \bibinfo {author} {\bibfnamefont {E.}~\bibnamefont {Yacoub}}, \bibinfo
  {author} {\bibfnamefont {K.}~\bibnamefont {Ugurbil}},\ and\ \bibinfo {author}
  {\bibnamefont {for~the WU-Minn HCP~Consortium}},\ }\bibfield  {title}
  {\bibinfo {title} {{The WU-Minn Human Connectome Project: An overview.}},\
  }\href@noop {} {\bibfield  {journal} {\bibinfo  {journal} {NeuroImage}\
  }\textbf {\bibinfo {volume} {80}},\ \bibinfo {pages} {62} (\bibinfo {year}
  {2013})}\BibitemShut {NoStop}%
\bibitem [{\citenamefont {Tuch}\ \emph {et~al.}(1999)\citenamefont {Tuch},
  \citenamefont {Wedeen}, \citenamefont {Dale}, \citenamefont {George},\ and\
  \citenamefont {Belliveau}}]{Tuch:1999}%
  \BibitemOpen
  \bibfield  {author} {\bibinfo {author} {\bibfnamefont {D.~S.}\ \bibnamefont
  {Tuch}}, \bibinfo {author} {\bibfnamefont {V.~J.}\ \bibnamefont {Wedeen}},
  \bibinfo {author} {\bibfnamefont {A.~M.}\ \bibnamefont {Dale}}, \bibinfo
  {author} {\bibfnamefont {J.~S.}\ \bibnamefont {George}},\ and\ \bibinfo
  {author} {\bibfnamefont {J.~W.}\ \bibnamefont {Belliveau}},\ }\bibfield
  {title} {\bibinfo {title} {{Conductivity Mapping of Biological Tissue Using
  Diffusion MRI}},\ }\href@noop {} {\bibfield  {journal} {\bibinfo  {journal}
  {Annals of the New York Academy of Sciences}\ }\textbf {\bibinfo {volume}
  {888}},\ \bibinfo {pages} {314} (\bibinfo {year} {1999})}\BibitemShut
  {NoStop}%
\bibitem [{\citenamefont {Haueisen}\ \emph {et~al.}(2002)\citenamefont
  {Haueisen}, \citenamefont {Tuch}, \citenamefont {Ramon}, \citenamefont
  {Schimpf}, \citenamefont {Wedeen}, \citenamefont {George},\ and\
  \citenamefont {Belliveau}}]{Haueisen:2002}%
  \BibitemOpen
  \bibfield  {author} {\bibinfo {author} {\bibfnamefont {J.}~\bibnamefont
  {Haueisen}}, \bibinfo {author} {\bibfnamefont {D.~S.}\ \bibnamefont {Tuch}},
  \bibinfo {author} {\bibfnamefont {C.}~\bibnamefont {Ramon}}, \bibinfo
  {author} {\bibfnamefont {P.~H.}\ \bibnamefont {Schimpf}}, \bibinfo {author}
  {\bibfnamefont {V.~J.}\ \bibnamefont {Wedeen}}, \bibinfo {author}
  {\bibfnamefont {J.~S.}\ \bibnamefont {George}},\ and\ \bibinfo {author}
  {\bibfnamefont {J.~W.}\ \bibnamefont {Belliveau}},\ }\bibfield  {title}
  {\bibinfo {title} {{The influence of brain tissue anisotropy on human EEG and
  MEG}},\ }\href@noop {} {\bibfield  {journal} {\bibinfo  {journal}
  {NeuroImage}\ }\textbf {\bibinfo {volume} {15}},\ \bibinfo {pages} {159}
  (\bibinfo {year} {2002})}\BibitemShut {NoStop}%
\bibitem [{\citenamefont {Hallez}\ \emph {et~al.}(2005)\citenamefont {Hallez},
  \citenamefont {Vanrumste}, \citenamefont {Hese}, \citenamefont {D'Asseler},
  \citenamefont {Lemahieu},\ and\ \citenamefont
  {de~Walle}}]{0031-9155-50-16-009}%
  \BibitemOpen
  \bibfield  {author} {\bibinfo {author} {\bibfnamefont {H.}~\bibnamefont
  {Hallez}}, \bibinfo {author} {\bibfnamefont {B.}~\bibnamefont {Vanrumste}},
  \bibinfo {author} {\bibfnamefont {P.~V.}\ \bibnamefont {Hese}}, \bibinfo
  {author} {\bibfnamefont {Y.}~\bibnamefont {D'Asseler}}, \bibinfo {author}
  {\bibfnamefont {I.}~\bibnamefont {Lemahieu}},\ and\ \bibinfo {author}
  {\bibfnamefont {R.~V.}\ \bibnamefont {de~Walle}},\ }\bibfield  {title}
  {\bibinfo {title} {A finite difference method with reciprocity used to
  incorporate anisotropy in electroencephalogram dipole source localization},\
  }\href {http://stacks.iop.org/0031-9155/50/i=16/a=009} {\bibfield  {journal}
  {\bibinfo  {journal} {Physics in Medicine \& Biology}\ }\textbf {\bibinfo
  {volume} {50}},\ \bibinfo {pages} {3787} (\bibinfo {year}
  {2005})}\BibitemShut {NoStop}%
\bibitem [{\citenamefont {Nunez}\ and\ \citenamefont
  {Srinivasan}(2014)}]{pmid24505628}%
  \BibitemOpen
  \bibfield  {author} {\bibinfo {author} {\bibfnamefont {P.~L.}\ \bibnamefont
  {Nunez}}\ and\ \bibinfo {author} {\bibfnamefont {R.}~\bibnamefont
  {Srinivasan}},\ }\bibfield  {title} {\bibinfo {title} {{{N}eocortical
  dynamics due to axon propagation delays in cortico-cortical fibers: {E}{E}{G}
  traveling and standing waves with implications for top-down influences on
  local networks and white matter disease}},\ }\href@noop {} {\bibfield
  {journal} {\bibinfo  {journal} {Brain Res.}\ }\textbf {\bibinfo {volume}
  {1542}},\ \bibinfo {pages} {138} (\bibinfo {year} {2014})}\BibitemShut
  {NoStop}%
\bibitem [{\citenamefont {Muller}\ \emph {et~al.}(2016)\citenamefont {Muller},
  \citenamefont {Piantoni}, \citenamefont {Koller}, \citenamefont {Cash},
  \citenamefont {Halgren},\ and\ \citenamefont {Sejnowski}}]{pmid27855061}%
  \BibitemOpen
  \bibfield  {author} {\bibinfo {author} {\bibfnamefont {L.}~\bibnamefont
  {Muller}}, \bibinfo {author} {\bibfnamefont {G.}~\bibnamefont {Piantoni}},
  \bibinfo {author} {\bibfnamefont {D.}~\bibnamefont {Koller}}, \bibinfo
  {author} {\bibfnamefont {S.~S.}\ \bibnamefont {Cash}}, \bibinfo {author}
  {\bibfnamefont {E.}~\bibnamefont {Halgren}},\ and\ \bibinfo {author}
  {\bibfnamefont {T.~J.}\ \bibnamefont {Sejnowski}},\ }\bibfield  {title}
  {\bibinfo {title} {{{R}otating waves during human sleep spindles organize
  global patterns of activity that repeat precisely through the night}},\
  }\href@noop {} {\bibfield  {journal} {\bibinfo  {journal} {Elife}\ }\textbf
  {\bibinfo {volume} {5}} (\bibinfo {year} {2016})}\BibitemShut {NoStop}%
\bibitem [{\citenamefont {Lubenov}\ and\ \citenamefont
  {Siapas}(2009)}]{pmid19489117}%
  \BibitemOpen
  \bibfield  {author} {\bibinfo {author} {\bibfnamefont {E.~V.}\ \bibnamefont
  {Lubenov}}\ and\ \bibinfo {author} {\bibfnamefont {A.~G.}\ \bibnamefont
  {Siapas}},\ }\bibfield  {title} {\bibinfo {title} {{{H}ippocampal theta
  oscillations are travelling waves}},\ }\href@noop {} {\bibfield  {journal}
  {\bibinfo  {journal} {Nature}\ }\textbf {\bibinfo {volume} {459}},\ \bibinfo
  {pages} {534} (\bibinfo {year} {2009})}\BibitemShut {NoStop}%
\bibitem [{\citenamefont {Zhang}\ \emph {et~al.}(2018)\citenamefont {Zhang},
  \citenamefont {Watrous}, \citenamefont {Patel},\ and\ \citenamefont
  {Jacobs}}]{pmid29887341}%
  \BibitemOpen
  \bibfield  {author} {\bibinfo {author} {\bibfnamefont {H.}~\bibnamefont
  {Zhang}}, \bibinfo {author} {\bibfnamefont {A.~J.}\ \bibnamefont {Watrous}},
  \bibinfo {author} {\bibfnamefont {A.}~\bibnamefont {Patel}},\ and\ \bibinfo
  {author} {\bibfnamefont {J.}~\bibnamefont {Jacobs}},\ }\bibfield  {title}
  {\bibinfo {title} {{{T}heta and {A}lpha {O}scillations {A}re {T}raveling
  {W}aves in the {H}uman {N}eocortex}},\ }\href@noop {} {\bibfield  {journal}
  {\bibinfo  {journal} {Neuron}\ }\textbf {\bibinfo {volume} {98}},\ \bibinfo
  {pages} {1269} (\bibinfo {year} {2018})}\BibitemShut {NoStop}%
\bibitem [{\citenamefont {Muller}\ \emph {et~al.}(2018)\citenamefont {Muller},
  \citenamefont {Chavane}, \citenamefont {Reynolds},\ and\ \citenamefont
  {Sejnowski}}]{pmid29563572}%
  \BibitemOpen
  \bibfield  {author} {\bibinfo {author} {\bibfnamefont {L.}~\bibnamefont
  {Muller}}, \bibinfo {author} {\bibfnamefont {F.}~\bibnamefont {Chavane}},
  \bibinfo {author} {\bibfnamefont {J.}~\bibnamefont {Reynolds}},\ and\
  \bibinfo {author} {\bibfnamefont {T.~J.}\ \bibnamefont {Sejnowski}},\
  }\bibfield  {title} {\bibinfo {title} {{{C}ortical travelling waves:
  mechanisms and computational principles}},\ }\href@noop {} {\bibfield
  {journal} {\bibinfo  {journal} {Nat. Rev. Neurosci.}\ }\textbf {\bibinfo
  {volume} {19}},\ \bibinfo {pages} {255} (\bibinfo {year} {2018})}\BibitemShut
  {NoStop}%
\bibitem [{\citenamefont {Fox}\ \emph {et~al.}(1986)\citenamefont {Fox},
  \citenamefont {Wolfson},\ and\ \citenamefont {Ranck}}]{pmid3720881}%
  \BibitemOpen
  \bibfield  {author} {\bibinfo {author} {\bibfnamefont {S.~E.}\ \bibnamefont
  {Fox}}, \bibinfo {author} {\bibfnamefont {S.}~\bibnamefont {Wolfson}},\ and\
  \bibinfo {author} {\bibfnamefont {J.~B.}\ \bibnamefont {Ranck}},\ }\bibfield
  {title} {\bibinfo {title} {{{H}ippocampal theta rhythm and the firing of
  neurons in walking and urethane anesthetized rats}},\ }\href@noop {}
  {\bibfield  {journal} {\bibinfo  {journal} {Exp Brain Res}\ }\textbf
  {\bibinfo {volume} {62}},\ \bibinfo {pages} {495} (\bibinfo {year}
  {1986})}\BibitemShut {NoStop}%
\bibitem [{\citenamefont {Stewart}\ \emph {et~al.}(1992)\citenamefont
  {Stewart}, \citenamefont {Quirk}, \citenamefont {Barry},\ and\ \citenamefont
  {Fox}}]{pmid1521610}%
  \BibitemOpen
  \bibfield  {author} {\bibinfo {author} {\bibfnamefont {M.}~\bibnamefont
  {Stewart}}, \bibinfo {author} {\bibfnamefont {G.~J.}\ \bibnamefont {Quirk}},
  \bibinfo {author} {\bibfnamefont {M.}~\bibnamefont {Barry}},\ and\ \bibinfo
  {author} {\bibfnamefont {S.~E.}\ \bibnamefont {Fox}},\ }\bibfield  {title}
  {\bibinfo {title} {{{F}iring relations of medial entorhinal neurons to the
  hippocampal theta rhythm in urethane anesthetized and walking rats}},\
  }\href@noop {} {\bibfield  {journal} {\bibinfo  {journal} {Exp Brain Res}\
  }\textbf {\bibinfo {volume} {90}},\ \bibinfo {pages} {21} (\bibinfo {year}
  {1992})}\BibitemShut {NoStop}%
\bibitem [{\citenamefont {Czurko}\ \emph {et~al.}(2011)\citenamefont {Czurko},
  \citenamefont {Huxter}, \citenamefont {Li}, \citenamefont {Hangya},\ and\
  \citenamefont {Muller}}]{pmid21414915}%
  \BibitemOpen
  \bibfield  {author} {\bibinfo {author} {\bibfnamefont {A.}~\bibnamefont
  {Czurko}}, \bibinfo {author} {\bibfnamefont {J.}~\bibnamefont {Huxter}},
  \bibinfo {author} {\bibfnamefont {Y.}~\bibnamefont {Li}}, \bibinfo {author}
  {\bibfnamefont {B.}~\bibnamefont {Hangya}},\ and\ \bibinfo {author}
  {\bibfnamefont {R.~U.}\ \bibnamefont {Muller}},\ }\bibfield  {title}
  {\bibinfo {title} {{{T}heta phase classification of interneurons in the
  hippocampal formation of freely moving rats}},\ }\href@noop {} {\bibfield
  {journal} {\bibinfo  {journal} {J. Neurosci.}\ }\textbf {\bibinfo {volume}
  {31}},\ \bibinfo {pages} {2938} (\bibinfo {year} {2011})}\BibitemShut
  {NoStop}%
\bibitem [{\citenamefont {Weiss}\ and\ \citenamefont
  {Faber}(2010)}]{pmid20508749}%
  \BibitemOpen
  \bibfield  {author} {\bibinfo {author} {\bibfnamefont {S.~A.}\ \bibnamefont
  {Weiss}}\ and\ \bibinfo {author} {\bibfnamefont {D.~S.}\ \bibnamefont
  {Faber}},\ }\bibfield  {title} {\bibinfo {title} {{{F}ield effects in the
  {C}{N}{S} play functional roles}},\ }\href@noop {} {\bibfield  {journal}
  {\bibinfo  {journal} {Front Neural Circuits}\ }\textbf {\bibinfo {volume}
  {4}},\ \bibinfo {pages} {15} (\bibinfo {year} {2010})}\BibitemShut {NoStop}%
\bibitem [{\citenamefont {Qiu}\ \emph {et~al.}(2015)\citenamefont {Qiu},
  \citenamefont {Shivacharan}, \citenamefont {Zhang},\ and\ \citenamefont
  {Durand}}]{pmid26631463}%
  \BibitemOpen
  \bibfield  {author} {\bibinfo {author} {\bibfnamefont {C.}~\bibnamefont
  {Qiu}}, \bibinfo {author} {\bibfnamefont {R.~S.}\ \bibnamefont
  {Shivacharan}}, \bibinfo {author} {\bibfnamefont {M.}~\bibnamefont {Zhang}},\
  and\ \bibinfo {author} {\bibfnamefont {D.~M.}\ \bibnamefont {Durand}},\
  }\bibfield  {title} {\bibinfo {title} {{{C}an {N}eural {A}ctivity {P}ropagate
  by {E}ndogenous {E}lectrical {F}ield?}},\ }\href@noop {} {\bibfield
  {journal} {\bibinfo  {journal} {J. Neurosci.}\ }\textbf {\bibinfo {volume}
  {35}},\ \bibinfo {pages} {15800} (\bibinfo {year} {2015})}\BibitemShut
  {NoStop}%
\bibitem [{\citenamefont {Zhang}\ \emph {et~al.}(2014)\citenamefont {Zhang},
  \citenamefont {Ladas}, \citenamefont {Qiu}, \citenamefont {Shivacharan},
  \citenamefont {Gonzalez-Reyes},\ and\ \citenamefont {Durand}}]{pmid24453330}%
  \BibitemOpen
  \bibfield  {author} {\bibinfo {author} {\bibfnamefont {M.}~\bibnamefont
  {Zhang}}, \bibinfo {author} {\bibfnamefont {T.~P.}\ \bibnamefont {Ladas}},
  \bibinfo {author} {\bibfnamefont {C.}~\bibnamefont {Qiu}}, \bibinfo {author}
  {\bibfnamefont {R.~S.}\ \bibnamefont {Shivacharan}}, \bibinfo {author}
  {\bibfnamefont {L.~E.}\ \bibnamefont {Gonzalez-Reyes}},\ and\ \bibinfo
  {author} {\bibfnamefont {D.~M.}\ \bibnamefont {Durand}},\ }\bibfield  {title}
  {\bibinfo {title} {{{P}ropagation of epileptiform activity can be independent
  of synaptic transmission, gap junctions, or diffusion and is consistent with
  electrical field transmission}},\ }\href@noop {} {\bibfield  {journal}
  {\bibinfo  {journal} {J. Neurosci.}\ }\textbf {\bibinfo {volume} {34}},\
  \bibinfo {pages} {1409} (\bibinfo {year} {2014})}\BibitemShut {NoStop}%
\bibitem [{\citenamefont {Chiang}\ \emph {et~al.}(2019)\citenamefont {Chiang},
  \citenamefont {Shivacharan}, \citenamefont {Wei}, \citenamefont
  {Gonzalez-Reyes},\ and\ \citenamefont {Durand}}]{pmid30295923}%
  \BibitemOpen
  \bibfield  {author} {\bibinfo {author} {\bibfnamefont {C.~C.}\ \bibnamefont
  {Chiang}}, \bibinfo {author} {\bibfnamefont {R.~S.}\ \bibnamefont
  {Shivacharan}}, \bibinfo {author} {\bibfnamefont {X.}~\bibnamefont {Wei}},
  \bibinfo {author} {\bibfnamefont {L.~E.}\ \bibnamefont {Gonzalez-Reyes}},\
  and\ \bibinfo {author} {\bibfnamefont {D.~M.}\ \bibnamefont {Durand}},\
  }\bibfield  {title} {\bibinfo {title} {{{S}low periodic activity in the
  longitudinal hippocampal slice can self-propagate non-synaptically by a
  mechanism consistent with ephaptic coupling}},\ }\href@noop {} {\bibfield
  {journal} {\bibinfo  {journal} {J. Physiol. (Lond.)}\ }\textbf {\bibinfo
  {volume} {597}},\ \bibinfo {pages} {249} (\bibinfo {year}
  {2019})}\BibitemShut {NoStop}%
\bibitem [{\citenamefont {Shivacharan}\ \emph {et~al.}(2019)\citenamefont
  {Shivacharan}, \citenamefont {Chiang}, \citenamefont {Zhang}, \citenamefont
  {Gonzalez-Reyes},\ and\ \citenamefont {Durand}}]{pmid30776338}%
  \BibitemOpen
  \bibfield  {author} {\bibinfo {author} {\bibfnamefont {R.~S.}\ \bibnamefont
  {Shivacharan}}, \bibinfo {author} {\bibfnamefont {C.~C.}\ \bibnamefont
  {Chiang}}, \bibinfo {author} {\bibfnamefont {M.}~\bibnamefont {Zhang}},
  \bibinfo {author} {\bibfnamefont {L.~E.}\ \bibnamefont {Gonzalez-Reyes}},\
  and\ \bibinfo {author} {\bibfnamefont {D.~M.}\ \bibnamefont {Durand}},\
  }\bibfield  {title} {\bibinfo {title} {{{S}elf-propagating, non-synaptic
  epileptiform activity propagates by endogenous electric fields}},\
  }\href@noop {} {\bibfield  {journal} {\bibinfo  {journal} {Exp. Neurol.}\ }
  (\bibinfo {year} {2019})}\BibitemShut {NoStop}%
\bibitem [{\citenamefont {Gabriel}\ \emph
  {et~al.}(1996{\natexlab{a}})\citenamefont {Gabriel}, \citenamefont {Lau},\
  and\ \citenamefont {Gabriel}}]{pmid8938025}%
  \BibitemOpen
  \bibfield  {author} {\bibinfo {author} {\bibfnamefont {S.}~\bibnamefont
  {Gabriel}}, \bibinfo {author} {\bibfnamefont {R.~W.}\ \bibnamefont {Lau}},\
  and\ \bibinfo {author} {\bibfnamefont {C.}~\bibnamefont {Gabriel}},\
  }\bibfield  {title} {\bibinfo {title} {{{T}he dielectric properties of
  biological tissues: {I}{I}. {M}easurements in the frequency range 10 {H}z to
  20 {G}{H}z}},\ }\href@noop {} {\bibfield  {journal} {\bibinfo  {journal}
  {Phys Med Biol}\ }\textbf {\bibinfo {volume} {41}},\ \bibinfo {pages} {2251}
  (\bibinfo {year} {1996}{\natexlab{a}})}\BibitemShut {NoStop}%
\bibitem [{\citenamefont {Gabriel}\ \emph
  {et~al.}(1996{\natexlab{b}})\citenamefont {Gabriel}, \citenamefont {Lau},\
  and\ \citenamefont {Gabriel}}]{pmid8938026}%
  \BibitemOpen
  \bibfield  {author} {\bibinfo {author} {\bibfnamefont {S.}~\bibnamefont
  {Gabriel}}, \bibinfo {author} {\bibfnamefont {R.~W.}\ \bibnamefont {Lau}},\
  and\ \bibinfo {author} {\bibfnamefont {C.}~\bibnamefont {Gabriel}},\
  }\bibfield  {title} {\bibinfo {title} {{{T}he dielectric properties of
  biological tissues: {I}{I}{I}. {P}arametric models for the dielectric
  spectrum of tissues}},\ }\href@noop {} {\bibfield  {journal} {\bibinfo
  {journal} {Phys Med Biol}\ }\textbf {\bibinfo {volume} {41}},\ \bibinfo
  {pages} {2271} (\bibinfo {year} {1996}{\natexlab{b}})}\BibitemShut {NoStop}%
\bibitem [{\citenamefont {Rayleigh}(1885)}]{citeulike:3581102}%
  \BibitemOpen
  \bibfield  {author} {\bibinfo {author} {\bibfnamefont {J.~W.~S.}\
  \bibnamefont {Rayleigh}},\ }\bibfield  {title} {\bibinfo {title} {{On Waves
  Propagated along the Plane Surface of an Elastic Solid}},\ }\href@noop {}
  {\bibfield  {journal} {\bibinfo  {journal} {Proceedings of the London
  Mathematical Society}\ }\textbf {\bibinfo {volume} {17}},\ \bibinfo {pages}
  {4} (\bibinfo {year} {1885})}\BibitemShut {NoStop}%
\bibitem [{\citenamefont {{Kartashov}}\ and\ \citenamefont
  {{Kuzelev}}(2014)}]{2014PlPhR..40..650K}%
  \BibitemOpen
  \bibfield  {author} {\bibinfo {author} {\bibfnamefont {I.~N.}\ \bibnamefont
  {{Kartashov}}}\ and\ \bibinfo {author} {\bibfnamefont {M.~V.}\ \bibnamefont
  {{Kuzelev}}},\ }\bibfield  {title} {\bibinfo {title} {{Dissipative surface
  waves in plasma}},\ }\href {https://doi.org/10.1134/S1063780X14070046}
  {\bibfield  {journal} {\bibinfo  {journal} {Plasma Physics Reports}\ }\textbf
  {\bibinfo {volume} {40}},\ \bibinfo {pages} {650} (\bibinfo {year}
  {2014})}\BibitemShut {NoStop}%
\bibitem [{\citenamefont {Galinsky}\ and\ \citenamefont
  {Frank}(2014)}]{Galinsky:2014}%
  \BibitemOpen
  \bibfield  {author} {\bibinfo {author} {\bibfnamefont {V.~L.}\ \bibnamefont
  {Galinsky}}\ and\ \bibinfo {author} {\bibfnamefont {L.~R.}\ \bibnamefont
  {Frank}},\ }\bibfield  {title} {\bibinfo {title} {Automated segmentation and
  shape characterization of volumetric data},\ }\href@noop {} {\bibfield
  {journal} {\bibinfo  {journal} {Neuroimage}\ }\textbf {\bibinfo {volume}
  {92}},\ \bibinfo {pages} {156} (\bibinfo {year} {2014})}\BibitemShut
  {NoStop}%
\bibitem [{\citenamefont {Galinsky}\ and\ \citenamefont
  {Frank}(2015)}]{Galinsky:2015}%
  \BibitemOpen
  \bibfield  {author} {\bibinfo {author} {\bibfnamefont {V.~L.}\ \bibnamefont
  {Galinsky}}\ and\ \bibinfo {author} {\bibfnamefont {L.~R.}\ \bibnamefont
  {Frank}},\ }\bibfield  {title} {\bibinfo {title} {Simultaneous multi-scale
  diffusion estimation and tractography guided by entropy spectrum pathways},\
  }\href@noop {} {\bibfield  {journal} {\bibinfo  {journal} {IEEE Trans. Med.
  Imag.}\ }\textbf {\bibinfo {volume} {34}},\ \bibinfo {pages} {1177} (\bibinfo
  {year} {2015})}\BibitemShut {NoStop}%
\bibitem [{\citenamefont {Galinsky}\ and\ \citenamefont
  {Frank}(2018{\natexlab{a}})}]{pmid30230014}%
  \BibitemOpen
  \bibfield  {author} {\bibinfo {author} {\bibfnamefont {V.~L.}\ \bibnamefont
  {Galinsky}}\ and\ \bibinfo {author} {\bibfnamefont {L.~R.}\ \bibnamefont
  {Frank}},\ }\bibfield  {title} {\bibinfo {title} {{{S}ymplectomorphic
  registration with phase space regularization by entropy spectrum pathways}},\
  }\href@noop {} {\bibfield  {journal} {\bibinfo  {journal} {Magn Reson Med}\
  ,\ \bibinfo {pages} {1}} (\bibinfo {year} {2018}{\natexlab{a}})}\BibitemShut
  {NoStop}%
\bibitem [{\citenamefont {Martinez}\ \emph {et~al.}(2015)\citenamefont
  {Martinez}, \citenamefont {Gaspar}, \citenamefont {Hillyard}, \citenamefont
  {Bickel}, \citenamefont {Lakatos}, \citenamefont {Dias},\ and\ \citenamefont
  {Javitt}}]{pmid26190988}%
  \BibitemOpen
  \bibfield  {author} {\bibinfo {author} {\bibfnamefont {A.}~\bibnamefont
  {Martinez}}, \bibinfo {author} {\bibfnamefont {P.~A.}\ \bibnamefont
  {Gaspar}}, \bibinfo {author} {\bibfnamefont {S.~A.}\ \bibnamefont
  {Hillyard}}, \bibinfo {author} {\bibfnamefont {S.}~\bibnamefont {Bickel}},
  \bibinfo {author} {\bibfnamefont {P.}~\bibnamefont {Lakatos}}, \bibinfo
  {author} {\bibfnamefont {E.~C.}\ \bibnamefont {Dias}},\ and\ \bibinfo
  {author} {\bibfnamefont {D.~C.}\ \bibnamefont {Javitt}},\ }\bibfield  {title}
  {\bibinfo {title} {{{N}eural oscillatory deficits in schizophrenia predict
  behavioral and neurocognitive impairments}},\ }\href@noop {} {\bibfield
  {journal} {\bibinfo  {journal} {Front Hum Neurosci}\ }\textbf {\bibinfo
  {volume} {9}},\ \bibinfo {pages} {371} (\bibinfo {year} {2015})}\BibitemShut
  {NoStop}%
\bibitem [{\citenamefont {Woldorff}\ \emph {et~al.}(2002)\citenamefont
  {Woldorff}, \citenamefont {Liotti}, \citenamefont {Seabolt}, \citenamefont
  {Busse}, \citenamefont {Lancaster},\ and\ \citenamefont
  {Fox}}]{pmid12433379}%
  \BibitemOpen
  \bibfield  {author} {\bibinfo {author} {\bibfnamefont {M.~G.}\ \bibnamefont
  {Woldorff}}, \bibinfo {author} {\bibfnamefont {M.}~\bibnamefont {Liotti}},
  \bibinfo {author} {\bibfnamefont {M.}~\bibnamefont {Seabolt}}, \bibinfo
  {author} {\bibfnamefont {L.}~\bibnamefont {Busse}}, \bibinfo {author}
  {\bibfnamefont {J.~L.}\ \bibnamefont {Lancaster}},\ and\ \bibinfo {author}
  {\bibfnamefont {P.~T.}\ \bibnamefont {Fox}},\ }\bibfield  {title} {\bibinfo
  {title} {{{T}he temporal dynamics of the effects in occipital cortex of
  visual-spatial selective attention}},\ }\href@noop {} {\bibfield  {journal}
  {\bibinfo  {journal} {Brain Res Cogn Brain Res}\ }\textbf {\bibinfo {volume}
  {15}},\ \bibinfo {pages} {1} (\bibinfo {year} {2002})}\BibitemShut {NoStop}%
\bibitem [{\citenamefont {Galinsky}\ and\ \citenamefont
  {Frank}(2018{\natexlab{b}})}]{FigShare}%
  \BibitemOpen
  \bibfield  {author} {\bibinfo {author} {\bibfnamefont {V.~L.}\ \bibnamefont
  {Galinsky}}\ and\ \bibinfo {author} {\bibfnamefont {L.~R.}\ \bibnamefont
  {Frank}},\ }\href@noop {} {\bibinfo {title} {{Brain Wave Loops movies}}}
  (\bibinfo {year} {2018}{\natexlab{b}}),\ \bibinfo {note} {supplemental
  Material on \href{https://figshare.com/s/08bda4716530cb9cfcd1}{FigShare} and
  \href{https://www.dropbox.com/sh/aads9hluqqzqfi6/AAC4Jzt8a-pH4FfJEf_6JUnBa?dl=0}{DropBox}}\BibitemShut
  {NoStop}%
\bibitem [{\citenamefont {Anastassiou}\ \emph {et~al.}(2011)\citenamefont
  {Anastassiou}, \citenamefont {Perin}, \citenamefont {Markram},\ and\
  \citenamefont {Koch}}]{pmid21240273}%
  \BibitemOpen
  \bibfield  {author} {\bibinfo {author} {\bibfnamefont {C.~A.}\ \bibnamefont
  {Anastassiou}}, \bibinfo {author} {\bibfnamefont {R.}~\bibnamefont {Perin}},
  \bibinfo {author} {\bibfnamefont {H.}~\bibnamefont {Markram}},\ and\ \bibinfo
  {author} {\bibfnamefont {C.}~\bibnamefont {Koch}},\ }\bibfield  {title}
  {\bibinfo {title} {{{E}phaptic coupling of cortical neurons}},\ }\href@noop
  {} {\bibfield  {journal} {\bibinfo  {journal} {Nat. Neurosci.}\ }\textbf
  {\bibinfo {volume} {14}},\ \bibinfo {pages} {217} (\bibinfo {year}
  {2011})}\BibitemShut {NoStop}%
\bibitem [{\citenamefont {Fonov}\ \emph {et~al.}(2011)\citenamefont {Fonov},
  \citenamefont {Evans}, \citenamefont {Botteron}, \citenamefont {Almli},
  \citenamefont {McKinstry}, \citenamefont {Collins}, \citenamefont {Ball},
  \citenamefont {Byars}, \citenamefont {Schapiro}, \citenamefont {Bommer},
  \citenamefont {Carr}, \citenamefont {German}, \citenamefont {Dunn},
  \citenamefont {Rivkin}, \citenamefont {Waber}, \citenamefont {Mulkern},
  \citenamefont {Vajapeyam}, \citenamefont {Chiverton}, \citenamefont {Davis},
  \citenamefont {Koo}, \citenamefont {Marmor}, \citenamefont {Mrakotsky},
  \citenamefont {Robertson}, \citenamefont {McAnulty}, \citenamefont {Brandt},
  \citenamefont {Fletcher}, \citenamefont {Kramer}, \citenamefont {Yang},
  \citenamefont {McCormack}, \citenamefont {Hebert}, \citenamefont {Volero},
  \citenamefont {Botteron}, \citenamefont {McKinstry}, \citenamefont {Warren},
  \citenamefont {Nishino}, \citenamefont {Almli}, \citenamefont {Todd},
  \citenamefont {Constantino}, \citenamefont {McCracken}, \citenamefont
  {Levitt}, \citenamefont {Alger}, \citenamefont {O'Neil}, \citenamefont
  {Toga}, \citenamefont {Asarnow}, \citenamefont {Fadale}, \citenamefont
  {Heinichen}, \citenamefont {Ireland}, \citenamefont {Wang}, \citenamefont
  {Moss}, \citenamefont {Zimmerman}, \citenamefont {Bintliff}, \citenamefont
  {Bradford}, \citenamefont {Newman}, \citenamefont {Evans}, \citenamefont
  {Arnaoutelis}, \citenamefont {Pike}, \citenamefont {Collins}, \citenamefont
  {Leonard}, \citenamefont {Paus}, \citenamefont {Zijdenbos}, \citenamefont
  {Das}, \citenamefont {Fonov}, \citenamefont {Fu}, \citenamefont {Harlap},
  \citenamefont {Leppert}, \citenamefont {Milovan}, \citenamefont {Vins},
  \citenamefont {Zeffiro}, \citenamefont {Van~Meter}, \citenamefont {Lange},
  \citenamefont {Froimowitz}, \citenamefont {Botteron}, \citenamefont {Almli},
  \citenamefont {Rainey}, \citenamefont {Henderson}, \citenamefont {Nishino},
  \citenamefont {Warren}, \citenamefont {Edwards}, \citenamefont {Dubois},
  \citenamefont {Smith}, \citenamefont {Singer}, \citenamefont {Wilber},
  \citenamefont {Pierpaoli}, \citenamefont {Basser}, \citenamefont {Chang},
  \citenamefont {Koay}, \citenamefont {Walker}, \citenamefont {Freund},
  \citenamefont {Rumsey}, \citenamefont {Baskir}, \citenamefont {Stanford},
  \citenamefont {Sirocco}, \citenamefont {Gwinn-Hardy}, \citenamefont
  {Spinella}, \citenamefont {McCracken}, \citenamefont {Alger}, \citenamefont
  {Levitt},\ and\ \citenamefont {O'Neill}}]{pmid20656036}%
  \BibitemOpen
  \bibfield  {author} {\bibinfo {author} {\bibfnamefont {V.}~\bibnamefont
  {Fonov}}, \bibinfo {author} {\bibfnamefont {A.~C.}\ \bibnamefont {Evans}},
  \bibinfo {author} {\bibfnamefont {K.}~\bibnamefont {Botteron}}, \bibinfo
  {author} {\bibfnamefont {C.~R.}\ \bibnamefont {Almli}}, \bibinfo {author}
  {\bibfnamefont {R.~C.}\ \bibnamefont {McKinstry}}, \bibinfo {author}
  {\bibfnamefont {D.~L.}\ \bibnamefont {Collins}}, \bibinfo {author}
  {\bibfnamefont {W.~S.}\ \bibnamefont {Ball}}, \bibinfo {author}
  {\bibfnamefont {A.~W.}\ \bibnamefont {Byars}}, \bibinfo {author}
  {\bibfnamefont {M.}~\bibnamefont {Schapiro}}, \bibinfo {author}
  {\bibfnamefont {W.}~\bibnamefont {Bommer}}, \bibinfo {author} {\bibfnamefont
  {A.}~\bibnamefont {Carr}}, \bibinfo {author} {\bibfnamefont {A.}~\bibnamefont
  {German}}, \bibinfo {author} {\bibfnamefont {S.}~\bibnamefont {Dunn}},
  \bibinfo {author} {\bibfnamefont {M.~J.}\ \bibnamefont {Rivkin}}, \bibinfo
  {author} {\bibfnamefont {D.}~\bibnamefont {Waber}}, \bibinfo {author}
  {\bibfnamefont {R.}~\bibnamefont {Mulkern}}, \bibinfo {author} {\bibfnamefont
  {S.}~\bibnamefont {Vajapeyam}}, \bibinfo {author} {\bibfnamefont
  {A.}~\bibnamefont {Chiverton}}, \bibinfo {author} {\bibfnamefont
  {P.}~\bibnamefont {Davis}}, \bibinfo {author} {\bibfnamefont
  {J.}~\bibnamefont {Koo}}, \bibinfo {author} {\bibfnamefont {J.}~\bibnamefont
  {Marmor}}, \bibinfo {author} {\bibfnamefont {C.}~\bibnamefont {Mrakotsky}},
  \bibinfo {author} {\bibfnamefont {R.}~\bibnamefont {Robertson}}, \bibinfo
  {author} {\bibfnamefont {G.}~\bibnamefont {McAnulty}}, \bibinfo {author}
  {\bibfnamefont {M.~E.}\ \bibnamefont {Brandt}}, \bibinfo {author}
  {\bibfnamefont {J.~M.}\ \bibnamefont {Fletcher}}, \bibinfo {author}
  {\bibfnamefont {L.~A.}\ \bibnamefont {Kramer}}, \bibinfo {author}
  {\bibfnamefont {G.}~\bibnamefont {Yang}}, \bibinfo {author} {\bibfnamefont
  {C.}~\bibnamefont {McCormack}}, \bibinfo {author} {\bibfnamefont {K.~M.}\
  \bibnamefont {Hebert}}, \bibinfo {author} {\bibfnamefont {H.}~\bibnamefont
  {Volero}}, \bibinfo {author} {\bibfnamefont {K.}~\bibnamefont {Botteron}},
  \bibinfo {author} {\bibfnamefont {R.~C.}\ \bibnamefont {McKinstry}}, \bibinfo
  {author} {\bibfnamefont {W.}~\bibnamefont {Warren}}, \bibinfo {author}
  {\bibfnamefont {T.}~\bibnamefont {Nishino}}, \bibinfo {author} {\bibfnamefont
  {C.~R.}\ \bibnamefont {Almli}}, \bibinfo {author} {\bibfnamefont
  {R.}~\bibnamefont {Todd}}, \bibinfo {author} {\bibfnamefont {J.}~\bibnamefont
  {Constantino}}, \bibinfo {author} {\bibfnamefont {J.~T.}\ \bibnamefont
  {McCracken}}, \bibinfo {author} {\bibfnamefont {J.}~\bibnamefont {Levitt}},
  \bibinfo {author} {\bibfnamefont {J.}~\bibnamefont {Alger}}, \bibinfo
  {author} {\bibfnamefont {J.}~\bibnamefont {O'Neil}}, \bibinfo {author}
  {\bibfnamefont {A.}~\bibnamefont {Toga}}, \bibinfo {author} {\bibfnamefont
  {R.}~\bibnamefont {Asarnow}}, \bibinfo {author} {\bibfnamefont
  {D.}~\bibnamefont {Fadale}}, \bibinfo {author} {\bibfnamefont
  {L.}~\bibnamefont {Heinichen}}, \bibinfo {author} {\bibfnamefont
  {C.}~\bibnamefont {Ireland}}, \bibinfo {author} {\bibfnamefont {D.~J.}\
  \bibnamefont {Wang}}, \bibinfo {author} {\bibfnamefont {E.}~\bibnamefont
  {Moss}}, \bibinfo {author} {\bibfnamefont {R.~A.}\ \bibnamefont {Zimmerman}},
  \bibinfo {author} {\bibfnamefont {B.}~\bibnamefont {Bintliff}}, \bibinfo
  {author} {\bibfnamefont {R.}~\bibnamefont {Bradford}}, \bibinfo {author}
  {\bibfnamefont {J.}~\bibnamefont {Newman}}, \bibinfo {author} {\bibfnamefont
  {A.~C.}\ \bibnamefont {Evans}}, \bibinfo {author} {\bibfnamefont
  {R.}~\bibnamefont {Arnaoutelis}}, \bibinfo {author} {\bibfnamefont {G.~B.}\
  \bibnamefont {Pike}}, \bibinfo {author} {\bibfnamefont {D.~L.}\ \bibnamefont
  {Collins}}, \bibinfo {author} {\bibfnamefont {G.}~\bibnamefont {Leonard}},
  \bibinfo {author} {\bibfnamefont {T.}~\bibnamefont {Paus}}, \bibinfo {author}
  {\bibfnamefont {A.}~\bibnamefont {Zijdenbos}}, \bibinfo {author}
  {\bibfnamefont {S.}~\bibnamefont {Das}}, \bibinfo {author} {\bibfnamefont
  {V.}~\bibnamefont {Fonov}}, \bibinfo {author} {\bibfnamefont
  {L.}~\bibnamefont {Fu}}, \bibinfo {author} {\bibfnamefont {J.}~\bibnamefont
  {Harlap}}, \bibinfo {author} {\bibfnamefont {I.}~\bibnamefont {Leppert}},
  \bibinfo {author} {\bibfnamefont {D.}~\bibnamefont {Milovan}}, \bibinfo
  {author} {\bibfnamefont {D.}~\bibnamefont {Vins}}, \bibinfo {author}
  {\bibfnamefont {T.}~\bibnamefont {Zeffiro}}, \bibinfo {author} {\bibfnamefont
  {J.}~\bibnamefont {Van~Meter}}, \bibinfo {author} {\bibfnamefont
  {N.}~\bibnamefont {Lange}}, \bibinfo {author} {\bibfnamefont {M.~P.}\
  \bibnamefont {Froimowitz}}, \bibinfo {author} {\bibfnamefont
  {K.}~\bibnamefont {Botteron}}, \bibinfo {author} {\bibfnamefont {C.~R.}\
  \bibnamefont {Almli}}, \bibinfo {author} {\bibfnamefont {C.}~\bibnamefont
  {Rainey}}, \bibinfo {author} {\bibfnamefont {S.}~\bibnamefont {Henderson}},
  \bibinfo {author} {\bibfnamefont {T.}~\bibnamefont {Nishino}}, \bibinfo
  {author} {\bibfnamefont {W.}~\bibnamefont {Warren}}, \bibinfo {author}
  {\bibfnamefont {J.~L.}\ \bibnamefont {Edwards}}, \bibinfo {author}
  {\bibfnamefont {D.}~\bibnamefont {Dubois}}, \bibinfo {author} {\bibfnamefont
  {K.}~\bibnamefont {Smith}}, \bibinfo {author} {\bibfnamefont
  {T.}~\bibnamefont {Singer}}, \bibinfo {author} {\bibfnamefont {A.~A.}\
  \bibnamefont {Wilber}}, \bibinfo {author} {\bibfnamefont {C.}~\bibnamefont
  {Pierpaoli}}, \bibinfo {author} {\bibfnamefont {P.~J.}\ \bibnamefont
  {Basser}}, \bibinfo {author} {\bibfnamefont {L.~C.}\ \bibnamefont {Chang}},
  \bibinfo {author} {\bibfnamefont {C.~G.}\ \bibnamefont {Koay}}, \bibinfo
  {author} {\bibfnamefont {L.}~\bibnamefont {Walker}}, \bibinfo {author}
  {\bibfnamefont {L.}~\bibnamefont {Freund}}, \bibinfo {author} {\bibfnamefont
  {J.}~\bibnamefont {Rumsey}}, \bibinfo {author} {\bibfnamefont
  {L.}~\bibnamefont {Baskir}}, \bibinfo {author} {\bibfnamefont
  {L.}~\bibnamefont {Stanford}}, \bibinfo {author} {\bibfnamefont
  {K.}~\bibnamefont {Sirocco}}, \bibinfo {author} {\bibfnamefont
  {K.}~\bibnamefont {Gwinn-Hardy}}, \bibinfo {author} {\bibfnamefont
  {G.}~\bibnamefont {Spinella}}, \bibinfo {author} {\bibfnamefont {J.~T.}\
  \bibnamefont {McCracken}}, \bibinfo {author} {\bibfnamefont {J.~R.}\
  \bibnamefont {Alger}}, \bibinfo {author} {\bibfnamefont {J.}~\bibnamefont
  {Levitt}},\ and\ \bibinfo {author} {\bibfnamefont {J.}~\bibnamefont
  {O'Neill}},\ }\bibfield  {title} {\bibinfo {title} {{{U}nbiased average
  age-appropriate atlases for pediatric studies}},\ }\href@noop {} {\bibfield
  {journal} {\bibinfo  {journal} {Neuroimage}\ }\textbf {\bibinfo {volume}
  {54}},\ \bibinfo {pages} {313} (\bibinfo {year} {2011})}\BibitemShut
  {NoStop}%
\bibitem [{\citenamefont {Fonov}\ \emph {et~al.}(2009)\citenamefont {Fonov},
  \citenamefont {Evans}, \citenamefont {McKinstry}, \citenamefont {Almli},\
  and\ \citenamefont {Collins}}]{FONOV2009S102}%
  \BibitemOpen
  \bibfield  {author} {\bibinfo {author} {\bibfnamefont {V.}~\bibnamefont
  {Fonov}}, \bibinfo {author} {\bibfnamefont {A.}~\bibnamefont {Evans}},
  \bibinfo {author} {\bibfnamefont {R.}~\bibnamefont {McKinstry}}, \bibinfo
  {author} {\bibfnamefont {C.}~\bibnamefont {Almli}},\ and\ \bibinfo {author}
  {\bibfnamefont {D.}~\bibnamefont {Collins}},\ }\bibfield  {title} {\bibinfo
  {title} {Unbiased nonlinear average age-appropriate brain templates from
  birth to adulthood},\ }\href
  {https://doi.org/https://doi.org/10.1016/S1053-8119(09)70884-5} {\bibfield
  {journal} {\bibinfo  {journal} {NeuroImage}\ }\textbf {\bibinfo {volume}
  {47}},\ \bibinfo {pages} {S102} (\bibinfo {year} {2009})},\ \bibinfo {note}
  {organization for Human Brain Mapping 2009 Annual Meeting}\BibitemShut
  {NoStop}%
\bibitem [{\citenamefont {Niethammer}\ \emph {et~al.}(2006)\citenamefont
  {Niethammer}, \citenamefont {San Jose~Estepar}, \citenamefont {Bouix},
  \citenamefont {Shenton},\ and\ \citenamefont {Westin}}]{pmid17946125}%
  \BibitemOpen
  \bibfield  {author} {\bibinfo {author} {\bibfnamefont {M.}~\bibnamefont
  {Niethammer}}, \bibinfo {author} {\bibfnamefont {R.}~\bibnamefont {San
  Jose~Estepar}}, \bibinfo {author} {\bibfnamefont {S.}~\bibnamefont {Bouix}},
  \bibinfo {author} {\bibfnamefont {M.}~\bibnamefont {Shenton}},\ and\ \bibinfo
  {author} {\bibfnamefont {C.~F.}\ \bibnamefont {Westin}},\ }\bibfield  {title}
  {\bibinfo {title} {{{O}n diffusion tensor estimation}},\ }\href@noop {}
  {\bibfield  {journal} {\bibinfo  {journal} {Conf Proc IEEE Eng Med Biol Soc}\
  }\textbf {\bibinfo {volume} {1}},\ \bibinfo {pages} {2622} (\bibinfo {year}
  {2006})}\BibitemShut {NoStop}%
\bibitem [{\citenamefont {Gomes}\ \emph {et~al.}(2016)\citenamefont {Gomes},
  \citenamefont {Bedard}, \citenamefont {Valtcheva}, \citenamefont {Nelson},
  \citenamefont {Khokhlova}, \citenamefont {Pouget}, \citenamefont {Venance},
  \citenamefont {Bal},\ and\ \citenamefont {Destexhe}}]{pmid26745426}%
  \BibitemOpen
  \bibfield  {author} {\bibinfo {author} {\bibfnamefont {J.~M.}\ \bibnamefont
  {Gomes}}, \bibinfo {author} {\bibfnamefont {C.}~\bibnamefont {Bedard}},
  \bibinfo {author} {\bibfnamefont {S.}~\bibnamefont {Valtcheva}}, \bibinfo
  {author} {\bibfnamefont {M.}~\bibnamefont {Nelson}}, \bibinfo {author}
  {\bibfnamefont {V.}~\bibnamefont {Khokhlova}}, \bibinfo {author}
  {\bibfnamefont {P.}~\bibnamefont {Pouget}}, \bibinfo {author} {\bibfnamefont
  {L.}~\bibnamefont {Venance}}, \bibinfo {author} {\bibfnamefont
  {T.}~\bibnamefont {Bal}},\ and\ \bibinfo {author} {\bibfnamefont
  {A.}~\bibnamefont {Destexhe}},\ }\bibfield  {title} {\bibinfo {title}
  {{{I}ntracellular {I}mpedance {M}easurements {R}eveal {N}on-ohmic
  {P}roperties of the {E}xtracellular {M}edium around {N}eurons}},\ }\href@noop
  {} {\bibfield  {journal} {\bibinfo  {journal} {Biophys. J.}\ }\textbf
  {\bibinfo {volume} {110}},\ \bibinfo {pages} {234} (\bibinfo {year}
  {2016})}\BibitemShut {NoStop}%
\bibitem [{\citenamefont {Barbour}(2017)}]{pmid28978453}%
  \BibitemOpen
  \bibfield  {author} {\bibinfo {author} {\bibfnamefont {B.}~\bibnamefont
  {Barbour}},\ }\bibfield  {title} {\bibinfo {title} {{{A}nalysis of {C}laims
  that the {B}rain {E}xtracellular {I}mpedance {I}s {H}igh and
  {N}on-resistive}},\ }\href@noop {} {\bibfield  {journal} {\bibinfo  {journal}
  {Biophys. J.}\ }\textbf {\bibinfo {volume} {113}},\ \bibinfo {pages} {1636}
  (\bibinfo {year} {2017})}\BibitemShut {NoStop}%
\bibitem [{\citenamefont {Galinsky}\ and\ \citenamefont
  {Frank}(2016)}]{brain-grid}%
  \BibitemOpen
  \bibfield  {author} {\bibinfo {author} {\bibfnamefont {V.~L.}\ \bibnamefont
  {Galinsky}}\ and\ \bibinfo {author} {\bibfnamefont {L.~R.}\ \bibnamefont
  {Frank}},\ }\bibfield  {title} {\bibinfo {title} {The lamellar structure of
  the brain fiber pathways},\ }\href@noop {} {\bibfield  {journal} {\bibinfo
  {journal} {Neural Comput.}\ }\textbf {\bibinfo {volume} {28}},\ \bibinfo
  {pages} {2533} (\bibinfo {year} {2016})}\BibitemShut {NoStop}%
\bibitem [{\citenamefont {Tuch}\ \emph {et~al.}(2001)\citenamefont {Tuch},
  \citenamefont {Wedeen}, \citenamefont {Dale}, \citenamefont {George},\ and\
  \citenamefont {Belliveau}}]{pmid11573005}%
  \BibitemOpen
  \bibfield  {author} {\bibinfo {author} {\bibfnamefont {D.~S.}\ \bibnamefont
  {Tuch}}, \bibinfo {author} {\bibfnamefont {V.~J.}\ \bibnamefont {Wedeen}},
  \bibinfo {author} {\bibfnamefont {A.~M.}\ \bibnamefont {Dale}}, \bibinfo
  {author} {\bibfnamefont {J.~S.}\ \bibnamefont {George}},\ and\ \bibinfo
  {author} {\bibfnamefont {J.~W.}\ \bibnamefont {Belliveau}},\ }\bibfield
  {title} {\bibinfo {title} {{{C}onductivity tensor mapping of the human brain
  using diffusion tensor {M}{R}{I}}},\ }\href@noop {} {\bibfield  {journal}
  {\bibinfo  {journal} {Proc. Natl. Acad. Sci. U.S.A.}\ }\textbf {\bibinfo
  {volume} {98}},\ \bibinfo {pages} {11697} (\bibinfo {year}
  {2001})}\BibitemShut {NoStop}%
\end{thebibliography}
%

\end{document}